\documentclass[10pt]{article}
\usepackage{amssymb}
\usepackage{amsmath}
\usepackage{amsthm}
\usepackage{latexsym}
\usepackage[dvips]{epsfig}
\usepackage{mathrsfs}
\usepackage{eufrak}

\theoremstyle{plain}
\newtheorem{proposition}{Proposition}
\newtheorem{lemma}{Lemma}
\newtheorem{theorem}{Theorem}

\newtheorem*{conjecture}{Conjecture}

\newtheorem{corollary}{Corollary}
\newtheorem*{main}{Theorem}
\newtheorem*{definition}{Definition}

\font\SYM=msbm10
\newcommand{\Real}{\mbox{\SYM R}}
\newcommand{\Complex}{\mbox{\SYM C}}

\newcommand{\Sphere}{\mbox{\SYM S}}

\font\tenscr=rsfs10 scaled1100
\font\sevenscr=rsfs7 % scaled \magstep1
\font\fivescr=rsfs5 % scaled \magstep1
\skewchar\tenscr='177
\skewchar\sevenscr='177
\skewchar\fivescr='177
\newfam\scrfam
\textfont\scrfam=\tenscr
\scriptfont\scrfam=\sevenscr
\scriptscriptfont\scrfam=\fivescr

\def\O{\mathcal{O}}

\def\pb{p_\bullet}

\newcommand{\TT}[3]{T_{#1 \phantom{#2} #3}^{\phantom{#1} #2}}
\newcommand{\updn}[3]{#1^{#2}_{\phantom{#2}#3}}
\newcommand{\dnup}[3]{#1_{#2}^{\phantom{#2}#3}}

\newcommand{\dnupdn}[4]{#1_{#2\phantom{#3}#4}^{\phantom{#2}#3}}

\begin{document}

% Luise
%\bibliographystyle{/home/jav/tex/reporthack}
% QM 1
%\bibliographystyle{/home/network/jav/tex/reporthack}
% Weyl
%\bibliographystyle{/home/jav/QM/tex/reporthack}
% Ludovica
%\bibliographystyle{/Users/Juan/Documents/tex/reporthack}

\title{\textbf{A rigidity property of asymptotically simple spacetimes arising from conformally flat data}}

\author{{\Large Juan Antonio Valiente Kroon} \thanks{E-mail address:
 {\tt j.a.valiente-kroon@qmul.ac.uk}} \\
School of Mathematical Sciences, Queen Mary, University of London,\\
Mile End Road, London E1 4NS, United Kingdom.}

\maketitle

\begin{abstract}
  Given a time symmetric initial data set for the vacuum Einstein
  field equations which is conformally flat near infinity, it is
  shown that the solutions to the regular finite initial value problem
  at spatial infinity extend smoothly through the critical sets where null
  infinity touches spatial infinity if and only if the initial data
  coincides with Schwarzschild data near infinity.
\end{abstract}


%PACS: 

\section{Introduction}

A vacuum spacetime
$(\tilde{\mathcal{M}},\tilde{g}_{\mu\nu})$ is said to be
\emph{asymptotically simple} ---cfr. \cite{Fri03a,Fri04,PenRin86}--- if there exists a smooth, oriented, time-oriented, causal spacetime
$(\mathcal{M},g_{\mu\nu})$ and a smooth function $\Xi$ (the \emph{conformal
factor}) on $\mathcal{M}$ such that:
\begin{itemize}
\item[(i)] $\mathcal{M}$ is a manifold with boundary $\mathscr{I}$, 

\item[(ii)] $\Xi>0$ on $\mathcal{M}\setminus \mathscr{I}$ and $\Xi=0$, $\mbox{d}\Xi \neq 0$ on $\mathscr{I}$, 

\item[(iii)] there exists a conformal embedding $\Phi:\tilde{\mathcal{M}}\rightarrow \mathcal{M}\setminus\mathscr{I}$ such that 
\[
\Xi^2\Phi^{-1*}\tilde{g}_{\mu\nu}=g_{\mu\nu},
\]

\item[(iv)] each null geodesic of
$(\tilde{\mathcal{M}},\tilde{g}_{\mu\nu})$ acquires two distinct
endpoints on $\mathscr{I}$.
\end{itemize}

\noindent
In this definition and throughout this article the word \emph{smooth}
is used as a synonym for $C^\infty$. The point (iv) in the definition
of asymptotic simplicity excludes black hole spacetimes. In order to
incorporate this class of spacetimes into the framework, the
definition has to be modified. A way of doing this is through the
notion of \emph{weak asymptotic simplicity}
---cfr. \cite{PenRin86,Ste91}. This point will not be of relevance in
our considerations.

\medskip
 The definition of asymptotic simplicity was introduced in
\cite{Pen63} as a way of characterising the asymptotic behaviour of
spacetimes describing isolated systems. As such, it allows to
reformulate questions concerning the asymptotic decay of physical
fields in terms of its differentiability at the conformal boundary. In
particular, asymptotic simplicity implies the peeling behaviour on the
Weyl tensor of the physical spacetime
$(\tilde{\mathcal{M}},\tilde{g}_{\mu\nu})$ ---see
e.g. \cite{PenRin86}. 

\medskip
It has been long debated whether the condition on the smoothness of
the conformal boundary is a too strong requirement to be imposed on a
spacetime describing interesting physical phenomena. In
\cite{ChrDel02} it has been shown that there exists a large class of
non-trivial asymptotically simple spacetimes. These spacetimes have
the peculiarity of being isometric to the Schwarzschild spacetime in a
neighbourhood of spatial infinity. Given this particular state of
affairs, the question is now: is this the only possible way to
construct non-trivial asymptotically simple spacetimes?

\medskip
The Cauchy problem in General Relativity provides a systematic
approach to address the existence of more general classes
of asymptotically simple spacetimes. In the spirit of
asymptotic simplicity, it is natural to work with objects and
equations which are defined in the conformally rescaled spacetime
$(\mathcal{M},g_{\mu\nu})$. The \emph{conformal Einstein field equations}
introduced in \cite{Fri81a,Fri81b,Fri82} and extensions thereof ---see
\cite{Fri95,Fri98a,Fri03a,Fri04}--- provide a suitable system of
equations and unknowns. These conformal field equations are formally
regular at the points where the conformal factor $\Xi$ vanishes, and
such that a solution of them implies a solution to the vacuum
Einstein field equations. In particular, the conformal Einstein field
equations have been used to prove a semi-global existence and stability
result for hyperboloidal data \cite{Fri86b} ---see also
\cite{LueVal09}. The construction of asymptotically simple spacetimes
in \cite{ChrDel02} makes use of this result.

\medskip
Cauchy data for asymptotically simple spacetimes is prescribed on
hypersurfaces which are asymptotically Euclidean. The compactification
of an asymptotically Euclidean hypersurface includes a singled out
point for each asymptotic end ---the corresponding ``point at
infinity''. These points become the \emph{spatial infinities} of the
development of the Cauchy data. Note however, that the definition of
asymptotic simplicity makes no reference to these points; the reason
being that if the spacetime has non-zero ADM mass then spatial
infinity is a singular point of the conformal geometry ---see
e.g. \cite{Fri88}.

\medskip
It has long been acknowledged that the main obstacle in the
construction of asymptotically simple spacetimes from Cauchy initial
data is the lack of a detailed understanding of the structure of the
Einstein field equations at spatial infinity
---cfr. \cite{Fri86a,Fri88}. The difficulties in the analysis of the
structure of spatial infinity arise from the singular nature of this
point with respect to the conformal geometry ---as mentioned in the
previous paragraph. In particular, the Weyl tensor is singular at
spatial infinity if the ADM mass is non-zero. Thus, one needs to
obtain a better representation of spatial infinity.

\medskip
The regular finite initial value problem
near spatial infinity introduced in \cite{Fri98a} provides a powerful
tool for analysing the properties of the gravitational field in the
regions of spacetime ``close'' to both spatial and null infinity. This
initial value problem makes use of the so-called \emph{extended conformal
Einstein field equations} and general properties of conformal
structures. It is such that both the equations and the data are
regular at the conformal boundary ---\emph{the regular finite initial
value problem at spatial infinity}.

\medskip
Whereas the standard compactification of spacetimes put forward in the
definition of asymptotically simple spacetimes considers spatial
infinity as a point, the approach used in \cite{Fri98a} represents
spatial infinity as an extended set with the topology $[-1,1]\times
\Sphere^2$. This so-called \emph{cylinder at spatial infinity} is
obtained as follows: starting from an asymptotically Euclidean initial
data set for the Einstein vacuum equations, one performs a
``standard'' conformal compactification to obtain a compact manifold,
$\mathcal{S}$, with a singled out point, $i$, representing the
infinity of the initial hypersurface ---for simplicity it is assumed
there is only one asymptotic end.  In a second stage, the point $i$ is
blown-up to a 2-sphere. This blowing-up of $i$ is achieved through the
lifting of a neighbourhood of $i$ to the bundle of space-spinor
dyads. In the final step, one uses a congruence of timelike conformal
geodesics to obtain Gaussian coordinates in a neighbourhood of the
initial hypersurface.  The conformal geodesics are conformal
invariants that can be used to construct a gauge which renders a
\emph{canonical} class of conformal factors for the development of the
initial data. These conformal factors can be written entirely in terms
of initial data quantities. Hence, the location of the conformal
boundary is known \emph{a priori}. The conformal boundary described by
the canonical conformal factor contains a null infinity with the usual
structure and a spatial infinity which extends in the time
dimension ---so that one can speak of the cylinder at spatial
infinity. The sets $\{\pm 1\} \times \Sphere^2$ will be called
\emph{critical sets} as they can be regarded as the collection of
points where null infinity ``touches'' spatial infinity. Null infinity
and spatial infinity do not meet tangentially at the critical
points. As a consequence, part of the propagation equations implied by
the conformal field equations degenerates at these sets. The analysis
in \cite{Fri98a} has shown that, as a result, the solutions to the
conformal field equations develop certain types of logarithmic
singularities at the critical sets. These singularities form an
intrinsic part of the conformal structure. It is to be expected that
the hyperbolic nature of the propagation equations will propagate
these logarithmic singularities. Consequently, they will have an
effect on the regularity of null infinity. However, there is not proof
of this yet.

\bigskip
The analysis carried out in \cite{Fri98a} ---restricted to
the developments of time symmetric data with an analytic
conformal compactification at infinity--- rendered an infinite set of
necessary conditions on the initial data for the spacetime to extend
analytically through the critical sets. These conditions can be
formulated in terms of the Cotton tensor of the initial metric
and its higher order derivatives. If these conditions are not
satisfied at some order then the solutions to the conformal Einstein
equations develop singularities of a very definite type at the
critical sets. It is important to point out that these singularities
are associated to structural properties of the principal part of the
evolution equations.

\bigskip 
The ideas and techniques of \cite{Fri98a} can be implemented
in a computer algebra system. In \cite{Val04a} this approach has been
used to analyse a certain type of asymptotic expansions of the
development of initial data sets which are conformally flat in a
neighbourhood of infinity. The rationale behind the use of conformally
flat data sets is that they satisfy the regularity conditions of
\cite{Fri98a} automatically. They constitute the simplest type of
initial data sets on which the methods of \cite{Fri98a} can be applied
to obtain non-trivial results. Although analytically simple,
conformally flat initial data sets give rise to spacetimes of full
complexity, and indeed, particular examples are used routinely in the
simulation of head-on black hole collisions ---see
e.g. \cite{AnnHobSei93,BriLin63,Mis63}. The key finding of
\cite{Val04a} was to show the existence of a further type of
logarithmic singularities at the critical sets. This new class of
obstructions to the smoothness of null infinity arises from the
interaction of the principal part (which is singular at the critical
sets) and the lower order terms in the conformal Einstein field
equations. An important observation arising from the analysis in
\cite{Val04a} is that particular subsets of these singularities can be
eliminated by setting certain pieces of the initial data to
zero. Proceeding in this way it is possible to gain insight into the
algebraic structure of the conformal field equations at spatial
infinity, and a very definite pattern is observed. One has the
following conjecture.

\begin{conjecture} \label{conjecture} 
  If an initial data set for the
  Einstein vacuum equations which is time symmetric and conformally
  flat in a neighbourhood of infinity yields a
  development with a smooth null infinity, then the initial data set
  is exactly Schwarzschildean in the neighbourhood.
\end{conjecture}

In this article we make definite progress towards a proof of this
conjecture. The analysis presented in this article is concerned with
the behaviour of the solutions of the conformal Einstein field
equations at the critical sets. The main result of this article is the
following:

\begin{main}
  Consider a time symmetric initial data set for the Einstein vacuum
  field equations which is conformally flat near infinity. The
  solution to the regular finite initial value problem at spatial
  infinity is smooth through the critical sets if and only if the data
  is exactly Schwarzschildean in a neighbourhood of infinity.
\end{main}

The result presented here falls short of providing a proof of the
conjecture for it is \emph{a priori} not clear that smoothness at null
infinity follows from the smoothness at the critical sets ---although
the expectation is that this will be the case. Conversely, the
connection between a singular behaviour at the critical sets and
non-smoothness at null infinity is, as yet, not fully understood. It
is expected that estimates of the type constructed in \cite{Fri03b}
should provide a way of linking the asymptotic expansions obtained by
the methods of \cite{Fri98a} with actual solutions to the conformal
field equations. A generalisation of these estimates to the setting
of a linear field propagating on a curved background has been
constructed in \cite{Val09}.

\bigskip
In order to bring our main result into a wider context, it is pointed
out that the Schwarzschild spacetime is the only static spacetime
admitting conformally flat slices ---see
e.g. \cite{Bei91b,Fri04}. Thus, our main result together with the
analysis in \cite{Val04d} of asymptotic expansions for more general
classes of time symmetric data suggests that it should be possible to
prove a generalisation in which smoothness at the critical sets
implies staticity in a neighbourhood of spatial infinity. The
situation for initial data sets with a non-vanishing second
fundamental form is not as clear cut as one loses the analyticity of
the conformal compactification ---cfr. \cite{DaiFri01}. In any case,
initial data sets which are stationary in a neighbourhood of infinity
are expected to play a privileged role ---see
e.g. \cite{Val04e,Val05a,Val06}.

\bigskip
The analysis leading to the main theorem provides very detailed
information about the behaviour of the solutions to the conformal
Einstein field at spatial infinity. In particular, it will be shown
that these singularities appear at higher orders than what it is to be
expected from general arguments. Indeed, there are some structural
properties of the conformal field equations (some cancellations) which
make the solutions more regular than what they are expected to be. This
explains why the first analysis carried out in \cite{FriKan00} did not
observe any singularities, and one had to undertake the higher order
expansions of \cite{Val04a}. It is certainly desirable to obtain a
deeper understanding of the structures responsible for this
behaviour. It is important to emphasise that these cancellations make
the analysis much more computationally challenging than what it
otherwise would be.

\subsection*{Outline of the article}

The present work is based on the analysis of \cite{Fri98a}
---cfr. also \cite{Fri04}. It makes use of a number of structures and
constructions which are not very standard. Therefore, in order to make
our discussion accessible, it will be necessary to introduce a certain
amount of notation and definitions. This should not be taken as a
comprehensive overview. The reader is, in any case, referred to
reference \cite{Fri98a} for complete details. I have striven, as much
as it is possible, to respect the original conventions of
\cite{Fri98a}. The main difference is the use of capital Latin letters
to denote spinorial indices.

\bigskip
Section \ref{section:conventions} introduces some basic notational
conventions and presents the notion of asymptotically Euclidean data
in the form that will be used here.

Section \ref{section:conformally:flat:data} provides a brief overview
of time symmetric, conformally flat initial data sets and introduces
the notion of data which is Schwarzschildean up to a certain order.

Section \ref{section:manifold_Ca} provides a summary of the blowing-up
of the point at spatial infinity. The blow-up is realised by the
introduction of a certain manifold $\mathcal{C}_a$ which is a subset
of the bundle of spin-frames. It is also shown how one can introduce a
calculus on this manifold, and the form that normal expansions acquire
when lifted up to this manifold. The contents of this section follow
reference \cite{Fri98a}, of which they constitute a very terse
summary.

Section \ref{section:Fgauge} discusses the extended conformal field
equations and the evolution equations they imply when written in a conformal
Gaussian gauge system ---the F-gauge. It discusses some structural
properties of these evolution equations and introduces the notion of
transport equations at the cylinder at spatial infinity.

Section \ref{section:Schwarzschild} discusses the Schwarzschild
spacetime in the F-gauge. This discussion will be a constant reference
point.

Section \ref{section:further:properties} presents the decomposition of
the linear transport equations into spherical harmonics. It introduces
the notion of a spherical harmonic sector and discusses the discrete
symmetries implied by the time reflection symmetry of the
spacetime. It also presents a procedure to solve the hierarchy of
transport equations. This procedure is different to the one given in
\cite{Fri98a} as it exploits a certain analogy with the Maxwell
equations ---cfr. \cite{Val07b}.

Section \ref{section:punchline} presents a systematic analysis of the
properties of the solutions of the transport equations for initial
data sets which are Schwarzschildean up to a certain order. It briefly reviews
 the computer algebra calculations of reference
\cite{Val04a}. It is shown that the solutions to these transport
equations are more regular than what one would \emph{a priori} expect,
but that neverthelesss, they eventually develop certain logarithmic
singularities at the critical sets. The order at which these
singularities arise is determined. The key results of this section are
based on very lengthy computer algebra calculations.Only qualitative
aspects of these calculations are presented.

In section \ref{section:main}, the results of the previous sections
are summarised in theorem \ref{thm:main}. This theorem is a more
precise version of the main result discussed in the introductory
paragraphs. Some further discussion is given, in particular regarding 
the potential generalisation to initial data sets which are not
conformally flat.

\section{Notation and conventions}
\label{section:conventions}
This article is concerned with the asymptotic properties of spacetimes
$(\tilde{\mathcal{M}},\tilde{g}_{\mu\nu})$ solving the Einstein vacuum field
equations
\begin{equation}
\tilde{R}_{\mu\nu}=0. \label{Einstein:eqns}
\end{equation}
The metric $\tilde{g}_{\mu\nu}$ will be assumed to have signature
$(+,-,-,-)$ and $\mu,\nu,\ldots$ are spacetime indices taking the
values $0,\ldots,3$ . The spacetime $(\tilde{\mathcal{M}},\tilde{g}_{\mu\nu})$
will be thought of as the development of some time symmetric initial
data prescribed on an asymptotically Euclidean Cauchy hypersurface
$\tilde{\mathcal{S}}$. The time symmetric data on
$\tilde{\mathcal{S}}$ are given in terms of a 3-metric
$\tilde{h}_{ij}$ of signature $(-,-,-)$. Due to the requirement of
time symmetry one has that the other piece of the data, the second
fundamental form of $\tilde{\mathcal{S}}$ vanishes: $\tilde{\chi}_{ij}=0$. The
letters $i,j,k,\ldots$ will be used as spatial tensorial indices
taking the values $1,2,3$. In this case the Einstein vacuum field
equations imply the constraint equation
\[
\tilde{r}=0
\]
where $\tilde{r}$ denotes 
the Ricci scalar of the metric $\tilde{h}_{ij}$. For simplicity, only
one asymptotically flat end will be assumed. In the asymptotically
flat end it will be assumed that coordinates $\{y^i\}$ can be
introduced such that
\[
\tilde{h}_{ij} =-\left(1+\frac{2m}{|y|}  \right) \delta_{ij} + \mathcal{O}\left( \frac{1}{|y|^2}\right), \quad \mbox{ as } |y|\rightarrow \infty,
\]
with $|y|^2 =(y^1)^2 +(y^2)^2 +(y^3)^2$ and $m$ a constant ---the ADM
mass of $\tilde{\mathcal{S}}$. In addition to these asymptotic
flatness requirements, it will be assumed that there is a
3-dimensional, orientable, smooth \emph{compact} manifold
$(\mathcal{S},h_{ij})$, a point $i \in \mathcal{S}$, a diffeomorphism
$\Phi:\mathcal{S}\setminus \{i\} \longrightarrow \tilde{\mathcal{S}}$
and a function $\Omega \in C^2(\mathcal{S}) \cap
C^\infty(\mathcal{S}\setminus \{i\})$ with the properties
\begin{subequations}
\begin{eqnarray}
&& \Omega(i)=0, \quad D_{j}\Omega(i)=0, \quad D_j D_k \Omega(i)=-2h_{jk}(i), \label{asymptotic:1} \\
&&  \Omega >0 \mbox{ on } \mathcal{S}\setminus \{ i\}, \label{asymptotic:2} \\
&& h_{ij} = \Omega^2 \Phi_* \tilde{h}_{ij}. \label{asymptotic:3}
\end{eqnarray}
\end{subequations}
The last condition shall be, sloppily, written as $h_{ij} = \Omega^2
\tilde{h}_{ij}$ ---that is, $\mathcal{S}\setminus \{i\}$ will be
identified with $\tilde{\mathcal{S}}$. Under these assumptions
$(\tilde{\mathcal{S}},\tilde{h}_{ij})$ will be said to be
\emph{asymptotically Euclidean and regular}. Suitable punctured
neighbourhoods of the point $i$ will be mapped into the asymptotic end
of $\tilde{\mathcal{S}}$. It should be clear from the context whether
$i$ denotes a point or a tensorial index.

\section{Conformally flat, time symmetric initial data sets}
\label{section:conformally:flat:data}

All throughout this article it will be assumed that one has initial
data sets which are conformally flat in a neighbourhood
$\mathcal{B}_a\subset \mathcal{S}$, $a>0$, of $i\in \mathcal{S}$.  The
expressions ``near infinity'' and ``in a neighbourhood of infinity''
used in the introductory paragraphs should be understood in this
sense.

\medskip
The Hamiltonian constraint together with the boundary conditions
(\ref{asymptotic:1})-(\ref{asymptotic:3}) imply on $\mathcal{B}_a(i)$
the \emph{Yamabe equation}
\begin{equation}
\label{Yamabe:equation}
\Delta \vartheta = -4\pi \delta(i), \quad \vartheta \equiv \Omega^{-2},
\end{equation}
where $\delta(i)$ denotes the Dirac delta distribution with support on
$i$, and $\Delta$ is the flat Laplacian. Let $x^{i}$ denote Cartesian
coordinates on $\mathcal{B}_a$ such that $x^i(i)=0$. On
$\mathcal{B}_a\setminus i$ the physical 3-metric $\tilde{h}_{ij}$ is
given by $\tilde{h}_{ij}=-\vartheta^4 \delta_{ij}$, where $\delta_{ij}$
is the standard Euclidean metric in Cartesian
coordinates. If $a>0$ is suitably small, the solutions of
(\ref{Yamabe:equation}) can be written on $\mathcal{B}_a$ in the
form
\begin{equation}
\label{theta_parametrised}
\vartheta = \frac{1}{|x|} + W,
\end{equation}
where $|x|=\sqrt{ (x^1)^2 + (x^2)^2 + (x^3)^2}$. The function $W$ satisfies
\begin{equation}
\label{Laplace:equation}
\Delta W =0, 
\end{equation}
so that $W$ is a harmonic and analytic function. From general
considerations on asymptotics, one has that
\[
W(i)=\frac{m}{2},
\]
where $m$ is the ADM mass of the time symmetric initial data set
$(\tilde{\mathcal{S}},\tilde{h}_{ij})$ ---cfr
\cite{Fri88,Fri98a,Fri04}. More generally, one has that
\begin{equation}
\label{Harmonic:function}
W= \frac{m}{2}+\sum^\infty_{k=1} w_{i_1 \cdots i_k} x^{i_1} \cdots x^{i_k}, 
\end{equation}
with $w_{i_1 \cdots i_k}$ denoting constant tensors which are totally
symmetric and $\delta$-tracefree so that the expression $w_{i_1 \cdots
i_k} x^{i_1} \cdots x^{i_k}$ is a homogeneous harmonic polynomial of
degree $k$. Conversely, due to the analyticity of the solutions to the
Laplace equation (\ref{Laplace:equation}), given a sequence
$\{w_{i_1\cdots i_k} \}$, $k=0,\ldots,\infty$, of constant totally
symmetric and trace-free constant tensors (the \emph{germ} of $W$)
such that the sum
\[
\sum^\infty_{k=1} w_{i_1 \cdots i_k} x^{i_1} \cdots x^{i_k},
\]
converges in $\mathcal{B}_a$ for a given $a>0$, then the sequence
defines a unique solution of the Yamabe equation
(\ref{Yamabe:equation}).

\bigskip
The following well-known observation will simplify our 
analysis. Consider the inversion $y^i=x^i/|x|^2$ mapping
$\mathcal{B}_a$ to the asymptotic end $\Real^3\setminus
\mathcal{B}_{1/a}(0)$, and let $\phi\equiv |x| \vartheta$. If
\[
\vartheta = \frac{1}{|x|} + \frac{m}{2} + w_i x^i + w_{ij} x^ix^j + \mathcal{O}(|x|^3),
\]
then it follows readily that
\[
\phi = 1 + \frac{m}{2|y|} + \frac{1}{|y|^3}w_i y^i + \frac{1}{|y|^5} w_{ij} y^iy^j + \mathcal{O}\left( \frac{1}{|y|^4}\right).
\]
Now, consider the translation $z^i=y^i+c^i$, with $c^i$ a constant
vector in $\Real^3$. A direct calculation shows that
\[
\frac{1}{|y|}=\frac{1}{|z|}\left(1-\frac{c_i z^i}{|z|^2} + \O\left( \frac{1}{|z|^2} \right)    \right), \quad \frac{1}{|y|^3}=\frac{1}{|z|^3}\left(1+ \O\left(\frac{1}{|z|}\right) \right),
\]
from where it follows that
\[
\phi = 1 + \frac{m}{2|z|} + \frac{1}{|z|^3} \left( w_i z^i - m c_i z^i   \right) + \O \left( \frac{1}{|z|^3} \right).
\]
Thus, the dipolar term can be removed by setting $c_i= w_i/m$
---that is, by choosing appropriately the ``centre of mass''. Finally,
an inversion of the form $z^i= \hat{x}^i/|\hat{x}|^2$, shows that one
can restrict the analysis to solutions to equation
(\ref{Yamabe:equation}) which in $\mathcal{B}_a(i)$ take the form
\[
\vartheta= \frac{1}{|x|} + \frac{m}{2} + w_{ij}x^ix^j + \O(|x|^3).
\]
In what follows it will always be assumed that $\vartheta$ is given in
the above form. In the above representation, data which is exactly
Schwarzschildean on $\mathcal{B}_a$ is characterised by
\[
W=\frac{m}{2}.
\]
More generally, we have the following.

\begin{definition}
  A time symmetric initial data set which is conformally flat in a
  suitable neighbourhood $\mathcal{B}_a$ of infinity will be said to
  be Schwarzschildean up to order $\pb$ if and only if
\[
\vartheta= \frac{1}{|x|} + \frac{m}{2} + w_{i_1 \cdots i_{p_\bullet+1}} x^{i_1} \cdots x^{i_{p_\bullet+1}} + \O ( |x|^{p_\bullet+2} ), 
\]
that is, if 
\[
W-m/2=\O ( |x|^{p_\bullet+1} ). 
\]
\end{definition}

\section{The Manifold $\mathcal{C}_a$} \label{section:manifold_Ca} 
In \cite{Fri98a} a representation of the region of spacetime close to
null infinity and spatial infinity has been introduced ---see also the
comprehensive discussion in \cite{Fri04}. The standard representation
of this region of spacetime depicts $i^0$ as a point.  In contrast,
the representation introduced in \cite{Fri98a} depicts spatial
infinity as a cylinder ---\emph{the cylinder at spatial
infinity}. This construction is briefly reviewed for the case of time
symmetric initial data sets which are conformally flat in a
neighbourhood $\mathcal{B}_a(i)$ of infinity. The reader is referred
to \cite{Fri98a,Fri04} for a thorough discussion of the details.

\subsection{The construction of $\mathcal{C}_a$}
Starting on the initial hypersurface $\mathcal{S}$, the construction
introduced in \cite{Fri98a} makes use of a blow-up of the
point $i\in \mathcal{S}$ to the 2-sphere $\Sphere^2$. This blow-up
requires the introduction of a particular bundle of spin-frames over
$\mathcal{B}_a$. Consider the (conformally rescaled)
spacetime $(\mathcal{M},g_{\mu\nu})$ obtained as the development of
the time symmetric initial data set $(\mathcal{S},h_{ij})$. Let
$SL(\mathcal{S})$ be the set of spin dyads
$\delta=\{\delta_A\}_{A=0,1}$ on $\mathcal{S}$ which are normalised
with respect to the alternating spinor $\epsilon_{AB}$ in such a way
that $\epsilon_{01}=1$. The set $SL(\mathcal{S})$ has a natural bundle
structure where $\mathcal{S}$ is the base space, and its structure
group is given by $SL(2,\Complex)$, acting on $SL(\mathcal{S})$ by
$\delta\mapsto \delta\cdot t=\{\delta_A
t^A_{\phantom{A}B}\}_{B=0,1}$. Let $\tau=\sqrt{2}e_0$, where $e_0$ is
the future $g$-unit normal of $\mathcal{S}$ and
\[
\tau_{AA'}=g(\tau,\delta_A\overline{\delta}_{A'})=\epsilon_{A}^{\phantom{A}0}\epsilon_{A'}^{\phantom{A'}0'}+\epsilon_{A}^{\phantom{A}1}\epsilon_{A'}^{\phantom{A'}1'}
\]
is its spinorial counterpart --- that is, $\tau=\tau^a e_a =
\sigma^a_{AA'}\tau^{AA'}e_a$ where $\sigma^a_{AA'}$ denote the
\emph{Infeld-van der Waerden symbols} and $\{e_a\}$, $a=0,\ldots,3$ is
an orthonormal frame. The spinor $\tau_{AA'}$ enables the introduction
of space-spinors ---sometimes also called $SU(2)$ spinors, see
\cite{Ash91,Fra98a,Som80}. It defines a sub-bundle $SU(\mathcal{S})$
of $SL(\mathcal{S})$ with structure group $SU(2,\Complex)$
and projection $\pi$ onto $\mathcal{S}$. The spinor $\tau^{AA'}$ allows to introduce \emph{spatial van der Waerden symbols} via
\[
\sigma_{a}^{AB}=\sigma^{(A}_{a\phantom{(A}A'}\tau^{B)A'}, \quad \sigma^a_{AB}=\tau_{(B}^{\phantom{(B}A'}\sigma^{a}_{\phantom{a}A)A'}, \quad i=1,2,3.
\]
The latter satisfy
\[
h_{ab}=\sigma_{aAB}\sigma_b^{AB}, \quad -\delta_{ab}\sigma^a_{AB}\sigma^b_{CD}=-\epsilon_{A(C}\epsilon_{D)B}\equiv h_{ABCD},
\]
with $h_{ab}=h(e_a,e_b)=-\delta_{ab}$. The bundle $SU(\mathcal{S})$ can be
endowed with a $\mathfrak{su}(2,\Complex)$-valued \emph{connection form}
$\check{\omega}^A_{\phantom{A}B}$ compatible with the metric
$h_{ij}$ and 1-form $\sigma^{AB}$, the \emph{solder form} of
$SU(S)$. The solder form satisfies by
construction
\begin{equation}
\label{metric}
h\equiv h_{ij} dx^i \otimes dx^j = h_{ABCD} \sigma^{AB}\otimes \sigma^{CD}.
\end{equation}

Given a spinorial dyad $\delta\in SU(\mathcal{S})$ one can define
an associated vector frame via $e_a=e_a(\delta)=\sigma^{AB}_a
\delta_A\tau_B^{\phantom{B}B'}\overline{\delta}_{B'}$, $a=1,2,3$.  We
shall restrict our attention to dyads related to frames
$\{e_j\}_{j=0,\cdots,3}$ on $\mathcal{B}_a$ such that $e_3$ is tangent
to the $h$-geodesics starting at $i$. Let $\check{H}$ denote the
horizontal vector field on $SU(\mathcal{S})$ projecting to the
radial vector $e_3$.

\medskip
The fibre $\pi^{-1}(i)\subset SU(\mathcal{S})$ (the fibre ``over'' $i$) can be
parametrised by choosing a fixed dyad $\delta^*$ and then letting the
group $SU(2,\Complex)$ act on it. Let $(-a,a)\ni \rho \mapsto
\delta(\rho,\updn{t}{A}{B})\in SU(\mathcal{S})$ be the integral curve to the vector
$\check{H}$ satisfying $\delta(0,\updn{t}{A}{B})=\delta(\updn{t}{A}{B})\in \pi^{-1}(i)$. With
this notation one defines the set
\[
\mathcal{C}_a =\big\{ \delta(\rho,\updn{t}{A}{B})\in SU(\mathcal{B}_{a}) \;\big|\; |\rho|<a, \; \updn{t}{A}{B}\in SU(2,\Complex)\big\},
\]
which is a smooth submanifold of $SU(\mathcal{S})$ diffeomorphic to
$(-a,a)\times SU(2,\Complex)$. The vector field $\check{H}$ is such
that its integral curves through the fibre $\pi^{-1}(i)$ project onto
the geodesics through $i$. From here it follows that the projection
map $\pi$ of the bundle $SU(\mathcal{S})$ maps $\mathcal{C}_a$ into
$\mathcal{B}_a$.

\medskip
Let, in the sequel, $\mathcal{I}^0\equiv \pi^{-1}(i)=\{\rho=0\}$ denote
the fibre over $i$. It can be seen that $\mathcal{I}^0\approx
SU(2,\Complex)$. On the other hand, for $p\in \mathcal{B}_a\setminus
\{i\}$ it turns out that $\pi^{-1}(p)$ consists of an orbit of $U(1)$
for which $\rho=|x(p)|$, and another for which $\rho=-|x(p)|$, where
$x^i(p)$ denote normal coordinates of the point $p$. In order to
understand better the structure of the manifold $\mathcal{C}_a$ it is
useful to quotient out the effect of $U(1)$. It turns out that
$\mathcal{I}^0/U(1)\approx \Sphere^2$. Hence, one has an extension of
the physical manifold $\tilde{S}$ by blowing up the point $i$ to
$\Sphere^2$.

\medskip
The manifold $\mathcal{C}_a$ inherits a number of structures from
$SU(\mathcal{S})$. In particular, the solder and connection forms can
be pulled back to smooth 1-forms on $\mathcal{C}_a$ satisfying the
\emph{structure equations} which relate them to the 
\emph{curvature form} determined by the \emph{curvature spinor}
$r_{ABCDEF}$. In the conformally flat setting one has that
\[
r_{ABCDEF}=0 \quad \mbox{ on } \mathcal{B}_a.
\]

\subsection{Calculus on $\mathcal{C}_a$}
In the sequel $\updn{t}{A}{B}\in SU(2,\Complex)$ and $\rho\in \Real$
will be used as coordinates on $\mathcal{C}_a$. Consequently, one has
that $\check{H}=\partial_\rho$. Vector fields relative to the
$SU(2,\Complex)$-dependent part of the coordinates are obtained by
looking at the basis of the (3-dimensional) Lie algebra
$\mathfrak{su}(2,\Complex)$ given by
\[
u_1\equiv\frac{1}{2}\left(
\begin{array}{cc}
0 & \mbox{i} \\
\mbox{i} & 0
\end{array}\right), \quad
u_2\equiv\frac{1}{2}\left(
\begin{array}{cc}
0 & -1 \\
1 & 0
\end{array}\right), \quad
u_3\equiv\frac{1}{2}\left(
\begin{array}{cc}
\mbox{i} & 0 \\
0 & -\mbox{i}
\end{array}\right).
\]
In particular, the vector $u_3$ is the generator of $U(1)$. Denote by
$Z_i$, $i=1,2,3$ the Killing vectors generated on $SU(\mathcal{S})$ by
$u_i$ and the action of $SU(2,\Complex)$. The vectors $Z_i$ are
tangent to $\mathcal{I}^0$. On $\mathcal{I}^0$ one sets
\begin{equation}
\label{diffops:X}
X_+=-(Z_2+\mbox{i}Z_1), \quad X_-=-(Z_2-\mbox{i}Z_1), \quad X=-2\mbox{i}Z_3,
\end{equation}
and extends these vector fields to the rest of $\mathcal{C}_a$ by
demanding them to commute with $\check{H}=\partial_\rho$. It is noted
that
\[
[X,X_+]=2X_+, \quad [X,X_-]=-2X_-, \quad [X_+,X_-]=-X.
\]
The vector fields are complex conjugates of each other in the sense
that for a given real-valued function $W$,
$\overline{X_-W}=X_+W$. More importantly, it can be seen that for
$p\in \mathcal{B}_a\setminus\{i\}$ the projections of the fields
$\partial_\rho$, $X_\pm$ span the tangent space at $p$.

\medskip
A frame $c_{AB}=c_{(AB)}$ dual to the solder forms
$\sigma^{CD}$ is defined so that it does not pick components along the fibres
---i.e. along the direction of $X$. These requirements imply
\begin{equation}
\label{solder_form}
\langle \sigma^{AB}, c_{CD} \rangle= h^{AB}_{\phantom{AB}CD}, \quad c_{CD}=c^1_{CD}\partial_\rho + c^+_{CD}X_+ + c^-_{CD}X_-,
\end{equation}
where $\langle \cdot , \cdot \rangle$ denotes the action of a 1-form
on a vector. In the conformally flat setting, from the properties of
the solder form $\sigma^{AB}$ one finds that
\begin{equation}
\label{frame_fields}
c^1_{AB}=x_{AB}, \quad c^+_{AB}=\frac{1}{\rho}z_{AB}, \quad c^-_{AB}=\frac{1}{\rho}y_{AB},
\end{equation}
with constant spinors $x_{AB}$, $y_{AB}$ and $z_{AB}$ given by
\[
x_{AB}\equiv \sqrt{2} \dnup{\epsilon}{(A}{0} \dnup{\epsilon}{B)}{1}, \quad y_{AB} \equiv -\frac{1}{\sqrt{2}}\dnup{\epsilon}{A}{1} \dnup{\epsilon}{B}{1}, \quad z_{AB}=\frac{1}{\sqrt{2}} \dnup{\epsilon}{A}{0} \dnup{\epsilon}{B}{0}.
\]
The connection coefficients, $\gamma_{ABCD}$, are defined by
contracting the connection form with the frame $c_{AB}$. In the
conformally flat case, one has
\begin{equation}
\label{conformally:flat:connection:coefficients}
\gamma_{ABCD} = \frac{1}{2\rho} (\epsilon_{AC}x_{BD}+\epsilon_{BD}x_{AC}).
\end{equation}
 
Let $f$ be a smooth function on $\mathcal{C}_a$
\[
D_{AB}f=c_{AB}(f).
\]
Similarly, let $\mu_{AB}$ be a spinor field on
$\mathcal{C}_a$. Then the covariant derivative of $\mu_{AB}$ is given
by
\[
D_{AB}\mu_{CD}= c_{AB}(\mu_{CD})- \gamma_{AB\phantom{E}C}^{\phantom{AB}E}\mu_{ED}-\gamma_{AB\phantom{E}D}^{\phantom{AB}E}\mu_{CE}.
\]
Analogous formulae hold for higher valence spinors.

\subsection{Normal expansions at $i$}
\label{subsection:normal_exp_i}
In \cite{Fri98a} a certain type of expansions of analytic fields near
$i$ has been discussed. Suppose $\xi^*_{A_1B_1\cdots A_lB_l}$ denotes
the components of an analytic even rank spinorial field with respect
to a fiduciary spin frame $\delta^*_A$. One can introduce the
expansion
\begin{equation}
\xi^*_{A_1B_1\cdots A_lB_l} (q)= \sum_{p=0}^\infty \frac{1}{p!}|x|^p n^{C_pD_p}\cdots n^{C_1D_1} D_{C_pD_p}\cdots D_{C_1D_1} \xi^*_{A_1B_1\cdots A_lB_l}(i), \label{spinor:expansion1}
\end{equation}
with $n^{AB}=n^{AB}(q)$, $q\in \mathcal{B}_a$, the spinorial
representation of the vector $n^i\partial_i=(x^i/|x|)\partial_i $, $n_i n^i=-1$.

\subsection{An orthonormal basis for functions on $SU(2,\Complex)$}
The lift of the expansion (\ref{spinor:expansion1}) from
$\mathcal{B}_a$ to $\mathcal{C}_a$ introduces in a natural way a class
of functions associated with unitary representations of
$SU(2,\Complex)$. Namely, given $\updn{t}{A}{B}\in SU(2,\Complex)$,
define
\begin{eqnarray*}
&& \TT{m}{j}{k}(\updn{t}{A}{B}) = \binom{m}{j}^{1/2} \binom{m}{k}^{1/2} \updn{t}{(B_1}{(A_1}\cdots \updn{t}{B_m)_j}{A_m)_k}, \\
&& \TT{0}{0}{0}(\updn{t}{A}{B})=1,
\end{eqnarray*}
with $j,k=0,\ldots,m$ and $m=1,2,3,\ldots$. The subindex
expression ${}_{(A_1\cdots A_m)_k}$ means that the indices are
symmetrised and then $k$ of them are set equal to $1$, while the
remaining ones are set to $0$. Details about the properties of these
functions can be found in \cite{Fri86a,Fri98a}.  The functions
$\sqrt{m+1}\TT{m}{j}{k}$ form a complete orthonormal set in the
Hilbert space $L^2(\mu,SU(2,\Complex))$, where $\mu$ denotes the
normalised Haar measure on $SU(2,\Complex)$. In particular, any
analytic complex-valued function $f$ on $SU(2,\Complex)$ admits an
expansion
\[
f(\updn{t}{A}{B})=\sum_{m=0}^\infty \sum_{j=0}^m \sum_{k=0}^m f_{m,k,j} \TT{m}{k}{j}(\updn{t}{A}{B}),
\]
with complex coefficients $f_{m,k,j}$. Under complex conjugation the
functions transform as
\begin{equation}
\label{complex_conjugation}
\overline{\TT{m}{j}{k}} =(-1)^{j+k} \TT{m}{m-j}{m-k}.
\end{equation}
The action of the differential operators (\ref{diffops:X}) on the
functions $\TT{m}{k}{j}$ is given by
\begin{eqnarray*}
&& X \TT{m}{k}{j}= (m-2j) \TT{m}{k}{j}, \\
&& X_+ \TT{m}{k}{j} = \sqrt{j(m-j+1)} \TT{m}{k}{j-1}, \quad X_-\TT{m}{k}{j}=-\sqrt{(j+1)(m-j)}\TT{m}{k}{j+1}.
\end{eqnarray*}
A function $f$ is said to have spin weight $s$ if
\[
Xf =2s f.
\]
Such a function has an expansion of the form
\[
f=\sum_{m\geq|2s|}^\infty \sum_{k=0}^m f_{m,k} \TT{m}{k}{m/2-s}.
\]
Finally it is noted that products $\TT{i_1}{j_1}{k_1}\times
\TT{i_2}{j_2}{k_2}$ can be \emph{linearised} ---that is, written as a linear
combination of other functions $\TT{i}{j}{k}$, for suitable $i$, $j$,
$k$ using the formula
\begin{eqnarray}
\label{Clebsch-Gordan}
&& \TT{i_1}{j_1}{k_1}\times \TT{i_2}{j_2}{k_2} = \sum^\mu_{p=0} D(i_1,j_1,k_1;i_2,j_2,k_2;i_1+i_2-2p,j_1+j_2-p,k_1+k_2-p) \nonumber \\
&&  \hspace{4cm}\times\TT{i_1+i_2-2p}{j_1+j_2-p}{k_1+k_2-p},
\end{eqnarray}
with $\mu=\min\{i_1,i_2,j_1+j_2,k_1+k_2\}$ and 
\[
D(i_1,j_1,k_1;i_2,j_2,k_2; l,m,n)= C(i_1,j_1;i_2,j_2;l,m)\overline{C(i_1,k_1;i_2,k_2;l,n)},
\]
and $C(i,j;k,l;m,n)$ the Clebsch-Gordan coefficients of $SU(2,\Complex)$.

\subsection{Normal expansions at $\mathcal{I}^0$}
\label{subsection:normal_exp_I0}
In the sequel, it will be necessary to \emph{lift} analytic fields
defined on $\mathcal{B}_a$ to $\mathcal{C}_a$. As in section
\ref{subsection:normal_exp_i}, consider normal coordinates $x^i$ on
$\mathcal{B}_a$ centred on $i$ and which are based on the orthonormal
frame $c^*_a = \sigma^{AB}_a c^*_{AB} =\sigma_a^{AB}\delta^*_A
\delta^*_B$. In terms of $\rho$ and $\updn{t}{A}{B}$ on
$\mathcal{C}_a$ and the normal coordinates $x^i$, the projection $\pi$
has the local expression
\[
\pi: (\rho,\updn{t}{A}{B}) \rightarrow x^i(\rho,\updn{t}{A}{B}) = \sqrt{2} \rho \sigma^i_{CD} \updn{t}{C}{0} \updn{t}{D}{1}.
\]
This expression can be used to carry out lifts to
$\mathcal{C}_a$. In particular, the lift of $|x|$ is
$\rho$. Applying the procedure described in \cite{Fri98a} to the
expansion (\ref{spinor:expansion1}) one obtains the
expansion of the spinor-valued function $\xi_{A_1B_1\cdots A_l B_l}$
on $\mathcal{C}_a$. Denote by $\xi_j=\xi_{(A_1B_1\cdots A_l B_l)_j}$,
$0\leq j \leq l$ its essential components. The function $\xi_j$ has
spin weight $s=l-j$ and a unique expansion of the form
\begin{equation}
\xi_j = \sum^\infty_{p=0} \xi_{j,p}\rho^p, \label{expansion:Ca1}
\end{equation}
with
\begin{equation}
\label{expansion:Ca2}
\xi_{j,p} = \sum_{q= \max\{|l-j|, l-p\}}^{p+l} \sum_{k=0}^{2q} \xi_{j,p;2q,k} \TT{2q}{k}{2q-l+j},
\end{equation}
and complex coefficients $\xi_{j,p;2q,k}$. More generally, we shall consider symmetric spinorial fields $\xi_{A_1\cdots
A_r}$ on $\mathcal{C}_a$ with independent
components $\xi_j = \xi_{(A_1\cdots A_{2r})_j}$, $0\leq j \leq 2r$,
and spin-weight $s=r-j$ which do not descend to analytic spinor fields
on $\mathcal{B}_a$. In this case one has that
\[
\xi_j =\sum^\infty_{p=0} \xi_{j,p}\rho^p, \quad \xi_{j,p}= \sum_{q=|r-j|}^{q(p)} \sum_{k=0}^{2q} \xi_{j,p;2q,k} \TT{2q}{k}{q-r+j},
\]
where one has \emph{a priori} that $0\leq |r-j| \leq q(p)$. An expansion of
the latter form will be said to be of type $q(p)$.

\subsubsection{On the expansions of $W$  on $\mathcal{C}_a$}
From the previous discussion it follows that the function $W$ on
$\mathcal{B}_a$ admits a lift to $\mathcal{C}_a$. This lift will be
again denoted by $W$. Its normal expansion at $\mathcal{I}^0$ is given by
\[
W= \frac{m}{2} + \sum_{p=2}^\infty \sum_{k=0}^{2p} \frac{1}{p!} w_{p;2p,k} \TT{2p}{k}{p} \rho^p,
\]
with $w_{p;2p,k}\in \Complex$ given by 
\begin{eqnarray*}
&& w_{p;2p,k}= (\sqrt{2})^p \binom{2p}{k}^{1/2} \binom{2p}{p}^{-1/2} D_{(B_pC_p} \cdots D_{B_1C_1)_k}W(i), \\
&& \phantom{ w_{p;2p,k}}=(\sqrt{2})^p \binom{2p}{k}^{1/2} \binom{2p}{p}^{-1/2} \sigma^{i_1}_{(B_1C_1} \cdots \sigma^{i_p}_{B_pC_p)_k} w_{i_1 \cdots i_p}.  
\end{eqnarray*}
Thus, $w_{i_1 \cdots i_p}=0$ if and only if $w_{p;p,k}=0$,
$k=0,\dots,2p$. The function $W$ is the lift to $\mathcal{C}_a$ of a
real function. Hence, it satisfies $W=\overline{W}$. Using the property
(\ref{complex_conjugation}) it follows that the coefficients
$w_{p;2p,k}$ satisfy the reality condition
\[
w_{p;2p,k}=(-1)^{p+k} \overline{w}_{p;2p,k}, \quad k=0,\dots,2p.
\] 
In particular, the coefficients $w_{p;2p,p}$ are real. If the function
$W$ is axially symmetric, then its normal expansions on
$\mathcal{C}_a$ take the simpler form
\[
W= \frac{m}{2} + \sum_{p=2}^\infty \frac{1}{p!} w_{p;2p,p} \TT{2p}{p}{p} \rho^p.
\]

\section{The spacetime Friedrich gauge}
\label{section:Fgauge}
 The formulation of the initial value problem near spatial infinity
presented in \cite{Fri98a} employs gauge conditions based on timelike
conformal geodesics. The conformal geodesics are curves which are
autoparallel with respect to a Weyl connection ---i.e. a torsion-free
connection which is not necessarily the Levi-Civita connection of a
metric. An analysis of Weyl connections in the context of the
conformal field equations has been given in \cite{Fri95}. In terms of
this gauge based on conformal geodesics ---which shall be called the
\emph{Friedrich gauge} or \emph{F-gauge} for short--- the conformal
factor of the spacetime can be determined explicitly in terms of the
initial data for the Einstein vacuum equations. Hence, provided that
the congruence of conformal geodesics and the fields describing the
gravitational field extend in a regular manner to null infinity, one
has complete control over the location of null infinity. This can be
ensured by making $\mathcal{B}_a$ suitably small. In addition, the
F-gauge renders a particularly simple representation of the
propagation equations. Using this framework, the singular initial
value problem at spatial infinity can be reformulated into another
problem where null infinity is represented by an explicitly known
hypersurface and where the data are regular at spacelike infinity.
The construction of the bundle manifold $\mathcal{C}_a$ and the
blowing up of the point $i\in \mathcal{B}_a$ to the set
$\mathcal{I}^0\subset \mathcal{C}_a$, briefly described in section
\ref{section:manifold_Ca}, are the first steps in the construction of
this regular setting. The next step is to introduce a rescaling of the
frame bundle so that fields that are singular at $\mathcal{I}^0$
become regular. This construction is briefly summarised in this
section.

\subsection{The  manifold $\mathcal{M}_{a,\kappa}$}
Following the discussion of \cite{Fri98a} assume that given the
development of data prescribed on $\mathcal{B}_a$, the timelike
spinor $\tau^{AA'}$ introduced in section
\ref{section:manifold_Ca} is tangent to a congruence of timelike
conformal geodesics which are orthogonal to $\mathcal{B}_a$. The
canonical factor rendered by this congruence
of conformal geodesics is given in terms of an affine parameter
$\tau$ of the conformal geodesics by
\begin{equation}
\Theta=\kappa^{-1}\Omega\left(1-\frac{\kappa^2\tau^2}{\omega^2}\right), \quad \mbox{ with } \quad  \omega=\frac{2\Omega}{\sqrt{|D_\alpha\Omega D^\alpha \Omega|}},\label{Theta}
\end{equation}
where $\Omega=\vartheta^{-2}$ and $\vartheta$ solves the Yamabe
equation (\ref{Yamabe:equation}) ---see
\cite{Fri95,Fri98a,Fri03c}. The function $\kappa>0$  expresses the remaining
conformal freedom in the construction. It will be
taken to be of the form $\kappa=\kappa^\prime \rho$, with
$\kappa^\prime$ analytic, $\kappa'(i)=1$. Associated to the conformal
factor $\Theta$ there is a 1-form $d_\mu$ from which the Weyl
connection can be obtained. In spinorial terms one has that
\[
d_{AA'}= \frac{1}{\sqrt{2}}\tau_{AA'} \partial_\tau{\Theta} -\updn{\tau}{B}{A'} d_{AB},
\]
where $d_{AB}$ is calculated in the case of  conformally flat data via
\[
d_{AB}= 2\rho \left( \frac{x_{AB}-\rho^2D_{AB}W}{(1+\rho W)^3} \right).
\]
The function $\kappa$ in the conformal factor $\Theta$, induces a
scaling $\delta_A \mapsto \kappa^{1/2} \delta_A$ of the spin
frame. Accordingly, one considers the bundle manifold
$\mathcal{C}_{a,\kappa}=\kappa^{1/2}\mathcal{C}_a$ of scaled spinor
frames. Using $\mathcal{C}_{a,\kappa}$ one defines the set 
\[
\mathcal{M}_{a,\kappa}=\left\{ (\tau,q) \big |  q\in
\mathcal{C}_{a,\kappa}, -\frac{\omega(q)}{\kappa(q)} \leq \tau \leq \frac{\omega(q)}{\kappa(q)} \right\},
\]
which, assuming that the congruence of null
geodesics and the relevant fields extend adequately, can be identified with the
development of $\mathcal{B}_a$ up to null infinity ---that is, the region of
spacetime near null and spatial infinity. In addition, one defines the sets:
\begin{subequations}
\begin{eqnarray*}
&& \mathcal{I}=\big \{(\tau,q)\in \mathcal{M}_{a,\kappa} \;\big|\; \rho(q)=0, \;|\tau|<1\big\}, \\
&& \mathcal{I}^\pm= \big \{ (\tau,q)\in \mathcal{M}_{a,\kappa} \;\big |\; \rho(q)=0, \;\tau=\pm1\big \}, \\
&& \mathscr{I}^\pm=\left\{ (\tau,q)\in \mathcal{M}_{a,\kappa} \;\big | \; \rho(q)>0, \;\; \tau=\pm \frac{\omega(q)}{\kappa(q)} \right\},
\end{eqnarray*}
\end{subequations}
which will be referred to as, respectively, the \emph{cylinder at
spatial infinity}, the \emph{critical sets} and \emph{future and past
null infinity}. In order to coordinatise the hypersurfaces of constant
parameter $\tau$, one extends the coordinates $(\rho,\updn{t}{A}{B})$
 off $\mathcal{C}_{a,\kappa}$ by requiring them to be constant along
the conformal geodesics ---i.e.  one has a system of \emph{conformal
Gaussian coordinates}.

\bigskip
\noindent
\textbf{Remark.} For the purposes of the analysis carried out in this
article it turns out that the most convenient choice of the function
$\kappa$ in the conformal factor $\Theta$ of equation (\ref{Theta}) is
\[
\kappa =\rho.
\]
This leads to considerable simplifications in all the relevant
expressions. From this point onwards, this choice will always be
assumed.

\subsection{The conformal propagation equations}
On the manifold $\mathcal{M}_{a,\kappa}$ it is possible to introduce a
calculus based on the derivatives $\partial_\tau$ and $\partial_\rho$
and on the operators $X_+$, $X_-$ and $X$. The operators
$\partial_\rho$, $X_+$, $X_-$ and $X$ originally defined on
$\mathcal{C}_{a}$ can be suitably extended to the rest of the manifold
by requiring them to commute with the vector field $\partial_\tau$. In
order to derive the propagation equations, a frame $c_{AA'}$ and the
associated spin connection coefficients $\Gamma_{AA'BC}$ of the Weyl
connection $\nabla$ will be used. The gravitational field is, in
addition, described by the spinorial counterparts of the Ricci tensor
of the Weyl connection, $\Theta_{AA'BB'}$, and of the rescaled Weyl
tensor, $\phi_{ABCD}$ ---see \cite{Fri95,Fri98a,Fri04}. Some further
notation will be required. Let
\[
\phi_{ABCD}= \phi_0 \epsilon^0_{ABCD} + \phi_1 \epsilon^1_{ABCD} + \phi_2 \epsilon^2_{ABCD} + \phi_3 \epsilon^3_{ABCD} + \phi_4 \epsilon^4_{ABCD},
\]
where
\[
\phi_i \equiv \phi_{(ABCD)_i}, \quad \epsilon^k_{ABCD}=\dnup{\epsilon}{(A}{(E} \dnup{\epsilon}{B}{F} \dnup{\epsilon}{C}{G} \dnup{\epsilon}{D)}{H)_i} \quad i=0,\ldots,4,
\]
where expressions like ${}_{(ABCD)_i}$ mean that after symmetrisation $i$
indices are set to 1. A space spinor $\Theta_{ABCD}=\Theta_{AB(CD)}$
is introduced such that
\[
\Theta_{AA'CC'}= \Theta_{ABCD} \updn{\tau}{B}{A'} \updn{\tau}{D}{C'}
\]
The space spinor $\Theta_{ABCD}$ shall be further decomposed as
\[
\Theta_{ABCD}=\Theta_{(AB)CD} + \frac{1}{2} \epsilon_{AB} \dnupdn{\Theta}{G}{G}{CD}.
\]
From the spin coefficients $\Gamma_{AA'BC}$ one defines
\[
\Gamma_{ABCD} \equiv \dnup{\tau}{B}{B'} \Gamma_{AB'CD},
\]
which in turn, will be decomposed as
\[
\Gamma_{ABCD} = \frac{1}{\sqrt{2}}\left(\xi_{ABCD} - \chi_{(AB)CD} \right) -\frac{1}{2}\epsilon_{AB} f_{CD}.
\]
They possess the following symmetries
\[
\Gamma_{ABCD}=\Gamma_{AB(CD)}, \quad \chi_{ABCD}=\chi_{AB(CD)}, \quad \xi_{ABCD}=\xi_{(AB)(CD)}. 
\]
The term $\xi_{ABCD}$ is related to the intrinsic connection of the leaves
of the foliation defined by $\tau_{AA'}$; the spinor $\chi_{(AB)CD}$
corresponds to the second fundamental form of the leaves; and $f_{AB}$
is the acceleration of the foliation.

\medskip
Using the F-gauge it can be shown that the extended conformal field
equations given in \cite{Fri98a} imply the following evolution
equations for the unknowns $c^\mu_{AB}$ ($\mu=0,1,\pm$), $\xi_{ABCD}$,
$f_{AB}$, $\chi_{(AB)CD}$, $\Theta_{(AB)CD}$,
$\Theta_{G\phantom{G}CD}^{\phantom{G}G}$:
\begin{subequations}
\begin{eqnarray}
&&\partial_\tau c^0_{AB}=-\dnup{\chi}{(AB)}{EF}c^{0}_{EF}-f_{AB}, \label{u:p1} \\
&&\partial_\tau c^\alpha_{AB}=-\dnup{\chi}{(AB)}{EF}c^\alpha_{EF}, \label{u:p2}\\
&&\partial_\tau \xi_{ABCD}=-\dnup{\chi}{(AB)}{EF}\xi_{EFCD}+\frac{1}{\sqrt{2}}(\epsilon_{AC}\chi_{(BD)EF}+\epsilon_{BD}\chi_{(AC)EF})f^{EF} \nonumber\\
&&\hspace{2cm} -\sqrt{2}\dnup{\chi}{(AB)(C}{E}f_{D)E}-\frac{1}{2}(\epsilon_{AC}\dnupdn{\Theta}{F}{F}{BD}+\epsilon_{BD}\dnupdn{\Theta}{F}{F}{AC})-\mbox{i}\Theta\mu_{ABCD}, \label{u:p3} \\
&&\partial_\tau f_{AB}=-\dnup{\chi}{(AB)}{EF}f_{EF}+\frac{1}{\sqrt{2}}\dnupdn{\Theta}{F}{F}{AB}, \label{u:p4} \\
&&\partial_\tau \chi_{(AB)CD}=-\dnup{\chi}{(AB)}{EF}\chi_{EFCD}-\Theta_{(CD)AB}+\Theta\eta_{ABCD}, \label{u:p5} \\
&&\partial_\tau\Theta_{(AB)CD}=-\dnup{\chi}{(CD)}{EF}\Theta_{(AB)EF}-\partial_\tau\Theta\eta_{ABCD}+\mbox{i}\sqrt{2}\updn{d}{E}{(A}\mu_{B)CDE}, \label{u:p6} \\
&&\partial_\tau \dnupdn{\Theta}{G}{G}{AB}=-\dnup{\chi}{(AB)}{EF}\dnupdn{\Theta}{G}{G}{EF}+\sqrt{2}d^{EF}\eta_{ABEF}, \label{u:p7}
\end{eqnarray}
\end{subequations}
where 
\begin{eqnarray*}
&&\eta_{ABCD}=\frac{1}{2}(\phi_{ABCD}+\dnup{\tau}{A}{A'}\dnup{\tau}{B}{B'}\dnup{\tau}{C}{C'}\dnup{\tau}{D}{D'} \overline{\phi}_{A'B'C'D'}), \\
&&\mu_{ABCD}=-\frac{\mbox{i}}{2}(\phi_{ABCD}-\dnup{\tau}{A}{A'}\dnup{\tau}{B}{B'}\dnup{\tau}{C}{C'}\dnup{\tau}{D}{D'}\overline{\phi}_{A'B'C'D'}),
\end{eqnarray*}
denote, respectively the electric and magnetic part of of
$\phi_{ABCD}$. Thus, the equations (\ref{u:p1})-(\ref{u:p7}) are
essentially ordinary differential equations for the components of
$c^\mu_{AB}$, $\Gamma_{ABCD}$, $\Theta_{ABCD}$. The redundancies in
the latter equations, which are due to the symmetries of the spinors,
can be eliminated by noting that valence-2 spinors can be
written as a linear combination of the spinors 
\[
x_{AB}\equiv \sqrt{2}\dnup{\epsilon}{(A}{0}\dnup{\epsilon}{B)}{1}, \quad y_{AB}\equiv -\frac{1}{\sqrt{2}}\dnup{\epsilon}{A}{1}\dnup{\epsilon}{B}{1}, \quad 
z_{AB}\equiv \frac{1}{\sqrt{2}} \dnup{\epsilon}{A}{0}\dnup{\epsilon}{B}{0}, 
\]
while the valence-4 spinors can be written in terms of the
spinors $\epsilon^i_{ABCD}$ ($i=0,\dots,4$), $h_{ABCD}$,
$x_{AC}\epsilon_{BD}+x_{BD}\epsilon_{AC}$,
$y_{AC}\epsilon_{BD}+y_{BD}\epsilon_{AC}$ and
$z_{AC}\epsilon_{BD}+z_{BD}\epsilon_{AC}$. The evolution equations for
the spinor $\phi_{ABCD}$ are derived from the Bianchi equations. One
has the following \emph{Bianchi propagation equations}:
\begin{subequations} 
\begin{eqnarray}
&&(\sqrt{2}-2c^0_{01})\partial_\tau\phi_0+2c^0_{00}\partial_\tau\phi_1-2c^\alpha_{01}\partial_{\alpha}\phi_0+2c^\alpha_{00}\partial_\alpha\phi_1 \nonumber \\
&&\hspace{1cm}= (2\Gamma_{0011}-8\Gamma_{1010})\phi_0+(4\Gamma_{0001}+8\Gamma_{1000})\phi_1-6\Gamma_{0000}\phi_2, \label{b0}\\
&&\sqrt{2}\partial_\tau\phi_1-c^0_{11}\partial_\tau\phi_0+c^0_{00}\partial_\tau\phi_2-c^\alpha_{11}\partial_{\alpha}\phi_0+c^\alpha_{00}\partial_{\alpha}\phi_2\nonumber\\
&&\hspace{1cm}=-(4\Gamma_{1110}+f_{11})\phi_0+(2\Gamma_{0011}+4\Gamma_{1100}-2f_{01})\phi_1
+3f_{00}\phi_2-2\Gamma_{0000}\phi_3, \label{b1} \\
&&\sqrt{2}\partial_\tau\phi_2-c^0_{11}\partial_\tau\phi_1+c^0_{00}\partial_\tau\phi_3-c^\alpha_{11}\partial_{\alpha}\phi_1+c^\alpha_{00}\partial_{\alpha}\phi_3\nonumber \\
&&\hspace{1cm}=-\Gamma_{1111}\phi_0-2(\Gamma_{1101}+f_{11})\phi_1+3(\Gamma_{0011}+\Gamma_{1100})\phi_2 \nonumber \\
&&\hspace{2cm}-2(\Gamma_{0001}-f_{00})\phi_3-\Gamma_{0000}\phi_4, \label{b2}\\
&&\sqrt{2}\partial_\tau\phi_3-c^0_{11}\partial_\tau\phi_2+c^0_{00}\partial_\tau\phi_4-c^\alpha_{11}\partial_{\alpha}\phi_2+c^\alpha_{00}\partial_{\alpha}\phi_4\nonumber \\
&&\hspace{1cm}=-2\Gamma_{1111}\phi_1
-3f_{11}\phi_2+(2\Gamma_{1100}+4\Gamma_{0011}+2f_{01})\phi_3-(4\Gamma_{0001}-f_{00})\phi_4,
\label{b3}\\
&&(\sqrt{2}+2c^0_{01})\partial_\tau\phi_4-2c^0_{11}\partial_\tau\phi_3+2c^\alpha_{01}\partial_\alpha\phi_4-2c^\alpha_{11}\partial_\alpha\phi_3 \nonumber \\
&&\hspace{1cm}=-6\Gamma_{1111}\phi_2+(4\Gamma_{1110}+8\Gamma_{0111})\phi_3
+(2\Gamma_{1100}-8\Gamma_{0101})\phi_4, \label{b4}
\end{eqnarray}
\end{subequations}
with $\alpha=1,\pm$ and $\partial_1\equiv \partial_\rho$,
$\partial_\pm\equiv X_\pm$. In addition to the latter propagation
equations, we shall also make use of a set of three equations, also
implied by the Bianchi identities, which we refer to as the
\emph{Bianchi constraint equations}:
\begin{subequations}  
\begin{eqnarray}
&&c^0_{11}\partial_\tau\phi_0-2c^0_{01}\partial_\tau\phi_1+c^0_{00}\partial_\tau \phi_2 + c^\alpha_{11}\partial_\alpha\phi_0-2c^\alpha_{01}\partial_\alpha\phi_1+c^\alpha_{00}\partial_\alpha\phi_2 \nonumber \\
&&\hspace{1cm}=-(2\Gamma_{(01)11}-4\Gamma_{1110})\phi_0+(2\Gamma_{0011}-4\Gamma_{(01)01}-4\Gamma_{1100})\phi_1 \nonumber \\
&& \hspace{2cm}+6\Gamma_{(01)00}\phi_2-2\Gamma_{0000}\phi_3, \label{bc1} \\
&&c^0_{11}\partial_\tau\phi_1-2c^0_{01}\partial_\tau\phi_2+c^0_{00}\partial_\tau \phi_3 + c^\alpha_{11}\partial_\alpha\phi_1-2c^\alpha_{01}\partial_\alpha\phi_2+c^\alpha_{00}\partial_\alpha\phi_3 \nonumber \\
&&\hspace{1cm}=\Gamma_{1111}\phi_0-(4\Gamma_{(01)11}-2\Gamma_{1101})\phi_1+3(\Gamma_{0011}-\Gamma_{1100})\phi_2 \nonumber \\
&&\hspace{2cm}-(2\Gamma_{0001}-4\Gamma_{(01)00})\phi_3-\Gamma_{0000}\phi_4, \label{bc2}\\
&&c^0_{11}\partial_\tau\phi_2-2c^0_{01}\partial_\tau\phi_3+c^0_{00}\partial_\tau \phi_4 + c^\alpha_{11}\partial_\alpha\phi_2-2c^\alpha_{01}\partial_\alpha\phi_3+c^\alpha_{00}\partial_\alpha\phi_4 \nonumber \\
&&\hspace{1cm}=2\Gamma_{1111}\phi_1-6\Gamma_{(01)11}\phi_2+(4\Gamma_{0011}+4\Gamma_{(01)01}-2\Gamma_{1100})\phi_3 \nonumber \\
&&\hspace{2cm}-(4\Gamma_{0001}-2\Gamma_{(01)00})\phi_4.
\label{bc3}
\end{eqnarray}
\end{subequations}

The propagation equations are supplemented by initial data on
$\mathcal{C}_{a,\kappa}$ constructed from the conformal factor
$\Omega=\vartheta^{-2}$ and the flat connection coefficients
$\gamma_{ABCD}$ ---cfr. equation
(\ref{conformally:flat:connection:coefficients})--- by using the
conformal constraint equations ---see e.g. \cite{Fri95} for more
details. One finds that for conformally flat data and the gauge choice $\kappa=\rho$:
\begin{subequations}
\begin{eqnarray}
&& \Theta_{ABCD}= -\frac{\rho^2}{\Omega}D_{(AB}D_{CD)}\Omega, \label{data1}\\
&& \phi_{ABCD}= \frac{\rho^3}{\Omega^2}D_{(AB}D_{CD)}\Omega, \label{data2}\\
&& c^0_{AB}=0, \label{data3a}\\ 
&& c^1_{AB}=\rho x_{AB}, \label{data3b}\\
&& c^+_{AB}= z_{AB}, \label{data4} \\
&& c^-_{AB}= y_{AB}, \label{data5}\\
&& \xi_{ABCD}=0, \label{data6}\\
&& \chi_{(AB)CD}=0, \label{data7} \\
&& f_{AB} = x_{AB} \label{data8},
\end{eqnarray}
\end{subequations}
on $\mathcal{C}_{a,\kappa}$. 

\subsection{Structural properties of the evolution equations}
We discuss now some general structural properties of the equations
(\ref{u:p1})-(\ref{u:p7}), (\ref{b1})-(\ref{b4}) and (\ref{bc1})-(\ref{bc3})
which will be used systematically in the sequel. Introduce the notation
\[
\upsilon \equiv \left(c^0_{AB}, c^\alpha_{AB}, \Gamma_{ABCD}, \Theta_{ABCD}\right), \quad \phi\equiv \left(\phi_0,\phi_1,\phi_2,\phi_3,\phi_4\right).
\]
The unknown vector $\upsilon$ has 45 independent complex components,
while $\phi$ has 5 independent complex components. In terms of $\upsilon$ and $\phi$, the
propagation equations (\ref{u:p1})-(\ref{u:p7}) can be written as:
\begin{equation}
\label{upsilon:propagation}
\partial_\tau \upsilon = K\upsilon + Q(\upsilon,\upsilon)+ L\phi,
\end{equation}
where $K$ and $Q$ denote, respectively, a linear and a quadratic
constant matrix-valued function with constant entries and $L$ is a linear
matrix-valued function with coefficients depending on the coordinates and
such that $L|_{\rho=0}=0$. Similarly, the systems (\ref{b1})-(\ref{b4})
and (\ref{bc1})-(\ref{bc3}) can be written as
\begin{subequations}
\begin{eqnarray}
&&\sqrt{2}E \partial_\tau \phi + A^{AB} c^\mu_{AB}\partial_\mu \phi =B(\Gamma_{ABCD})\phi, \label{bianchi:propagation}\\
&&F^{AB}c^\mu_{AB}\partial_{\mu}\phi =H(\Gamma_{ABCD}), \label{bianchi:constraint}
\end{eqnarray}
\end{subequations}
where $E$ denotes the $5\times 5$ unit matrix and $A^{AB}c^\mu_{AB}$,
$\mu=0,\ldots,3$, are $5\times 5$ matrices depending on the
coordinates, while $B(\Gamma_{ABCD})$ denotes a constant matrix-valued
linear function of the connection coefficients $\Gamma_{ABCD}$. On the
other hand, $F^{AB}c^\mu_{AB}$ denote $3\times 5$ matrices with
coordinate dependent entries and $H(\Gamma_{ABCD})$ is another 
constant matrix-valued linear function of the connection coefficients
$\Gamma_{ABCD}$.

\bigskip 
Consider now the system
(\ref{upsilon:propagation})-(\ref{bianchi:propagation}) with data
given on $\mathcal{C}_{a,\kappa}$. Given a neighbourhood
$\mathcal{W}$ of $\mathcal{C}_{a,\kappa}$ in $\mathcal{M}_{a,\kappa}$
on which a unique smooth solution of the Cauchy problem is
given. From the point of view of the propagation equations, the subset
$\mathcal{W}\cap \mathcal{I}$ is a regular hypersurface. Introduce the notation
$\upsilon^{(0)}\equiv \upsilon|_{\mathcal{W}\cap \mathcal{I}}$,
$\phi^{(0)} \equiv \phi|_{\mathcal{W}\cap \mathcal{I}}$. Due to the property 
$L|_{\rho=0}=0$, equations (\ref{upsilon:propagation}) decouple
from equations (\ref{bianchi:propagation}) and can be integrated
on $\mathcal{W}\cap \mathcal{I}$. Using the observation that the restriction of the initial data to $\mathcal{I}^0$ coincides with Minkowski data one has that on $\mathcal{W}\cap \mathcal{I}$:
\begin{subequations}
\begin{eqnarray}
&& \Theta^{(0)}=0, \quad \chi^{(0)}_{(AB)CD}=0, \quad f^{(0)}_{AB}=x_{AB}, \quad \xi^{(0)}_{ABCD}=0, \label{zeroth:order1}\\
&& (c^0_{AB})^{(0)}=-\tau x_{AB}, \quad (c^1_{AB})^{(0)}=0, \quad (c^-_{AB})^{(0)}=y_{AB}, \quad (c^+_{AB})^{(0)}=z_{AB}. \label{zeroth:order2}
\end{eqnarray}
\end{subequations}
It follows that $A^1\equiv A^{AB}c^1_{AB}$ is such that
\[
A^1|_{\mathcal{W}\cap \mathcal{I}}=0,
\]
so that the system (\ref{bianchi:propagation}) also implies an interior system on $\mathcal{W}\cap \mathcal{I}$ whose solution is
\begin{equation}
\phi^{(0)}_{ABCD} =-6m \epsilon^2_{ABCD}. \label{zeroth:order3}
\end{equation}
The solutions
(\ref{zeroth:order1})-(\ref{zeroth:order2}) and (\ref{zeroth:order3})
extend analytically to the whole of $\mathcal{I}$ and in particular to
the critical sets $\mathcal{I}^\pm$. The set $\mathcal{I}$ is a
\emph{total characteristic} of the system
(\ref{upsilon:propagation})-(\ref{bianchi:propagation}) in the sense
that the whole system reduces to an interior system on $\mathcal{I}$.
Moreover, the constraint equations (\ref{bianchi:constraint}) also
reduce to an interior system on $\mathcal{I}$. Another crucial
structural property is that 
\[
A^0\equiv \sqrt{2}E + A^{AB}c^0_{AB}= \sqrt{2}\mbox{diag}(1+\tau,1,1,1,1-\tau) \quad \mbox{ on } \mathcal{I},
\]
so that the matrix $A^0$ which is positive definite degenerates at
$\mathcal{I}^\pm$. Understanding the effects of this degeneracy is the
main motivation behind the analysis in this article.

\medskip
The previous discussion can be generalised by repeated application of
the differential operator $\partial_\rho$ to the equations
(\ref{upsilon:propagation}), (\ref{bianchi:propagation}) and
(\ref{bianchi:constraint}) to obtain interior systems for the
quantities $\upsilon^{(p)}=\partial^{(p)}_\rho
\upsilon|_{\mathcal{I}}$ and $\phi^{(p)}=\partial^{(p)}_\rho
\phi|_{\mathcal{I}}$ which will be called the \emph{order} $p$
\emph{transport equations}. Their behaviour on the whole of
$\mathcal{I}$ will be studied in the sequel. The transport equations
then take the following form for $p\geq 1$:
\begin{subequations}
\begin{eqnarray}
&&\partial_\tau v^{(p)} = Kv^{(p)}+Q(v^{(0)},v^{(p)})+Q(v^{(p)},v^{(0)}) \nonumber \\
&& \hspace{3cm}+\sum_{j=1}^{p-1}\binom{p}{j}\left(Q(v^{(j)},v^{(p-j)})+ L^{(j)}\phi^{(p-j)}\right) +
L^{(p)}\phi^{(0)}, \label{upsilon:transport} \\
&&\big(\sqrt{2}E+A^{AB}(c^0_{AB})^{(0)}\big)\partial_\tau\phi^{(p)} +
A^{AB}(c^\mu_{AB})^{(0)}\partial_\mu\phi^{(p)}= B(\Gamma^{(0)}_{ABCD})\phi^{(p)} \nonumber\\
&& \hspace{3cm} +\sum_{j=1}^p
\binom{p}{j}\left(B(\Gamma_{ABCD}^{(j)})\phi^{(p-j)}-A^{AB}(c^\mu_{AB})^{(j)}\partial_\mu
 \phi^{(p-j)}\right), \label{pbianchi:transport} \\
&&F^{AB}(c^0_{AB})^{(0)}\partial_\tau \phi^{(p)} +F^{AB}(c^\mu_{AB})^{(0)}\partial_\mu\phi^{(p)} =H(\Gamma^{(0)}_{ABCD})\phi^{(p)} \nonumber \\
&&\hspace{3cm} +\sum_{j=1}^p\binom{p}{j}\left(H(\Gamma^{(j)}_{ABCD})\phi^{(p-j)}-F^{AB}(c^\mu_{AB})^{(j)}\partial_\mu \phi^{(p-j)}\right). \label{cbianchi:transport}
\end{eqnarray}
\end{subequations}
In the previous equations the values of $\upsilon^{(0)}$ and
$\phi^{(0)}$ given in (\ref{zeroth:order1})-(\ref{zeroth:order2}) and
(\ref{zeroth:order3}) are assumed. Note that the non-homogeneous terms
in the equations (\ref{upsilon:transport})-(\ref{cbianchi:transport})
depend on $\upsilon^{(p')}$, $\phi^{(p')}$ for $0\leq p' <p$. Thus, if
their values are known, then
(\ref{upsilon:transport})-(\ref{pbianchi:transport}) constitutes an
interior system of linear equations for $\upsilon^{(p)}$ and
$\phi^{(p)}$. The principal part of these equations is universal, in
the sense that it is independent of the value of $p$. If the
initial data on $\mathcal{C}_{a,\kappa}$ for the system
(\ref{upsilon:propagation})-(\ref{bianchi:propagation}) is analytic
---as it is the case in the present analysis--- then suitable initial
data for the interior system
(\ref{upsilon:transport})-(\ref{pbianchi:transport}) can be obtained
by repeated $\rho$-differentiation and evaluation on $\mathcal{I}^0$.

\medskip
The interior system
(\ref{upsilon:transport})-(\ref{pbianchi:transport}) is decoupled in
the following sense: if $\upsilon^{(p')}$, $\phi^{(p')}$ are known for
$0\leq p' <p$ one can solve first (\ref{upsilon:transport}) as it contains at most quantities of order $\phi^{(p-1)}$. With the
knowledge of $\upsilon^{(p)}$ at hand one can then solve
(\ref{pbianchi:transport}) to obtain $\phi^{(p)}$.

\bigskip
The language of jets is natural in the present context. For
$p=0,1,2,\ldots$ and any sufficiently smooth (possibly vector valued)
function $f$ defined on $\mathcal{M}_{a,\kappa}$, the sets of
functions $\{f^{(0)}$, $f^{(1)}$, \ldots, $f^{(p)}\}$ on $\mathcal{I}$
will be denoted by $J^{(p)}_\mathcal{I}[f]$ and referred to as
\emph{the jet order $p$ of $f$ on} $\mathcal{I}$ ---and similarly with
$\mathcal{I}$ replaced by $\mathcal{I}^0$. If $u=(\upsilon,\phi)$ is a
solution to the equations (\ref{upsilon:transport}),
(\ref{pbianchi:transport}) and (\ref{cbianchi:transport}), we refer to
$J^{(p)}_\mathcal{I}[u]$ as to \emph{the s-jet of $u$ of order $p$}
and to the data $J^{(p)}_{\mathcal{I}^0}[u]$ as to \emph{the d-jet of
$u$ of order $p$}. An s-jet $J^{(p)}_\mathcal{I}[u]$ of order $p$ will
be called \emph{regular} on $ \overline{\mathcal{I}}\equiv \mathcal{I} \cup
\mathcal{I}^+ \cup \mathcal{I}^-$ if the corresponding functions
extend smoothly to the critical sets $\mathcal{I}^\pm$.

\section{The Schwarzschild spacetime in the F-gauge}
\label{section:Schwarzschild}

Due to its relevance for our purposes, a brief discussion of
the Schwarzschild spacetime in the F-gauge is provided. The material
is adapted from the one given in \cite{Fri98a}. The
Schwarzschild line element with mass $m$ in isotropic coordinates is
given by
\[
\tilde{g} = \left( \frac{1-m/2r}{1+m/2r}\right)^2 \mbox{d}t^2 -\left(1+\frac{m}{2r}\right)^4 \left( \mbox{d}r^2 +r^2 \mbox{d}\sigma^2  \right),
\]
where $\mbox{d}\sigma^2=\mbox{d}\theta^2 + \sin^2 \theta \mbox{d} \varphi^2$ is
the standard line element of the unit sphere $\Sphere^2$ in polar
coordinates. Writing the first fundamental form $\tilde{h}_{ij}$ and the second fundamental form $\tilde{\chi}_{ij}$ on the initial hypersurface $\tilde{S}=\{t=0\}$ in terms of the coordinate $\rho=1/r$ one obtains
\[
\tilde{h}_{ij}= \Omega^{-2} h_{ij}, \quad \tilde{\chi}_{ij}=0,
\]
where
\[
h =h_{ij}\mbox{d}x^i \mbox{d}x^j= -\left(\mbox{d}\rho^2 + \rho^2 \mbox{d}\sigma^2  \right), \quad \Omega=\frac{\rho^2}{(1+m\rho/2)^2}.
\]
Thus, we have initial data set for the Schwarzschild spacetime which is
time symmetric and conformally flat. Accordingly, one has
that
\[
U=1, \quad W=\frac{m}{2},
\]
near $\rho=0$. In what follows let $\kappa=\rho$ and assume that $a$
is chosen small enough such that $(1+(m/2)\rho)\neq 0$ for
$|\rho|<a$. The conformal factor $\Theta$ and the 1-form $d_{AB}$
associated to the F-gauge read in this case
\[
\Theta=\frac{\rho}{(1+\rho m/2)^2}\left(1-\frac{\tau^2}{(1+\rho m/2)^2} \right), \quad d_{AB}= \frac{2\rho x_{AB}}{(1+\rho m/2)^3}.
\]
Furthermore, the non-trivial initial data on $\mathcal{C}_{a,\kappa}$ is given by
\begin{eqnarray*}
&& \Theta_{(AB)CD}= \frac{6m \rho}{(1 + \rho m/2)^2} \epsilon^2_{ABCD}, \quad \dnupdn{\Theta}{G}{G}{CD}=0, \\
&& \phi_{ABCD}=-6m \epsilon^2_{ABCD}. 
\end{eqnarray*}
This initial data set for the propagation equations
(\ref{u:p1})-(\ref{u:p7}) and (\ref{b0})-(\ref{b4}) is explicitly
spherically symmetric ---the functions involved are of spin-weight 0
and contain only the harmonic $\TT{0}{0}{0}$.  Accordingly, one puts
forward a spherically symmetric Ansatz for their development. More
precisely, one writes
\begin{eqnarray*}
&& c^0_{AB} = c^0_x x_{AB}, \quad c^1_{AB}= c^1_x x_{AB}, \quad c^-_{AB}=c^-_y y_{AB}, \quad c^+_{AB}= c^+_z z_{AB}, \\
&& f_{AB}=f_x x_{AB}, \quad \xi_{ABCD}= \xi_x (\epsilon_{AC}x_{BD}+\epsilon_{BD}\epsilon_{AC}), \\
&&\chi_{(AB)CD}= \chi_2 \epsilon^2_{ABCD} + \chi_h h_{ABCD}, \\
&&\Theta_{(AB)CD}= \Theta_2 \epsilon^2_{ABCD} + \Theta_h h_{ABCD}, \quad \dnupdn{\Theta}{G}{G}{CD}= \theta_x x_{AB}, \\
&&\phi_{ABCD}= \phi_2 \epsilon^2_{ABCD},
\end{eqnarray*}
where
\[
c^0_x, \quad c^1_x, \quad c^+_z, \quad c^-_y, \quad f_x, \quad \xi_x, \quad \xi_2, \quad \chi_h, \quad \Theta_2, \quad \Theta_h, \quad \theta_x, \quad \phi_2,
\]
are functions depending on $(\rho,\tau)$ only. The previous Ansatz,
together with the propagation equations (\ref{u:p1})-(\ref{u:p7}) and
(\ref{b0})-(\ref{b4}) imply an initial value problem of the type
\begin{equation}
\label{Schwarzschild:equation}
\partial_\tau u = F(u,\tau,\rho;m), \quad u(0,\rho;m)= u_0(\rho;m),
\end{equation}
with analytic functions $F$ and $u_0$ for the unknowns 
\[
u=\left(c^0_x,c^1_x,c^-_y,c^+_z,f_x, \chi_x, \xi_2, \xi_h, \Theta_2, \Theta_h, \theta_x, \phi_2  \right). 
\]
The solution with $m=0$ corresponds to a portion of the conformal
Minkowski spacetime in which the only non-vanishing components of the
solution are given by
\[
c^0_x=-\tau, \quad c^1_x=\rho, \quad c^+_z=c^-_y=1, \quad f_x=1.
\]
Since in this case the solution exists for all $\tau, \;\rho\in
\Real$, it follows from standard results of ordinary differential
equations that for a given $m$ there is a sufficiently small $\rho_0$
such that there exists an analytic solution to the system
(\ref{Schwarzschild:equation}) which extends beyond $\mathscr{I}$ for
$|\rho|<|\rho_0$. Hence, one can recover the portion of the
Schwarzschild spacetime which lies near null and spatial infinity if $a$
is taken to be small enough. 

\medskip
It follows from the above discussion that the coefficients that are
obtained from solving the transport propagation equations on the
cylinder at spatial infinity 
correspond to the terms in the Taylor-like expansions
\[
u=\sum^\infty_{p=0} \frac{1}{p!} u^{(p)} \rho^p,
\]
of the solutions of the initial value problem
(\ref{Schwarzschild:equation}). In particular, the logarithmic
singularities that have been observed to appear in the critical sets
$\mathcal{I}^\pm$ of the development of more general classes of data
---see e.g. \cite{Val04a}--- do not arise in the case of the
Schwarzschild spacetime\footnote{More generally, this has been shown
to be the case for all asymptotically flat vacuum static spacetimes ---see
\cite{Fri04}.}. One has the following result.

\begin{proposition}
The solutions of the transport equations
(\ref{upsilon:transport}) and (\ref{pbianchi:transport}) for  time
symmetric Schwarzschild initial data extend analytically
through $\mathcal{I}^\pm$ for all orders $p$. Moreover, the solutions
to the transport equations are polynomial in $\tau$.
\end{proposition}

For the purposes of the present article it turns out that it will be necessary
to know the expansions explicitly up to order $p=4$ (inclusive). These
straightforward, but nevertheless lengthy computations have been
performed with the aid of a computer algebra system {\tt Maple V}. 

\section{Further properties of the Bianchi transport equations}
\label{section:further:properties}

The transport propagation equations for the Bianchi subsystem of
interior equations (\ref{pbianchi:transport}) read explicitly
\begin{subequations}
\begin{eqnarray}
&& (1+\tau) \partial_\tau \phi^{(p)}_0 + X_+\phi_1^{(p)} -(p-2) \phi_0^{(p)} = R^{(p)}_0, \label{expanded:pbianchi:transport:0}\\
&& \partial_\tau \phi_1^{(p)} +\frac{1}{2} X_+ \phi_2^{(p)} +\frac{1}{2} X_-\phi_0^{(p)} + \phi_1^{(p)} = R^{(p)}_1, \label{expanded:pbianchi:transport:1}\\
&& \partial_\tau \phi_2^{(p)} +\frac{1}{2} X_+\phi_3^{(p)} +\frac{1}{2}X_-\phi^{(p)}_1 = R^{(p)}_2, \label{expanded:pbianchi:transport:2}\\
&& \partial_\tau \phi^{(p)}_3 + \frac{1}{2}X_+ \phi^{(p)}_4 + \frac{1}{2} X_-\phi^{(p)}_2 -\phi^{(p)}_3 = R^{(p)}_3, \label{expanded:pbianchi:transport:3}\\
&& (1-\tau) \partial_\tau \phi^{(p)}_4 + X_-\phi^{(p)}_3 +(p-2)\phi^{(p)}_4 = R^{(p)}_4,\label{expanded:pbianchi:transport:4}
\end{eqnarray}
\end{subequations}
with $R_j=R_j(u^{(0)},\ldots,u^{(p-1)})$, $j=0,\ldots,4$. On the other
hand, the Bianchi transport constraint equations
(\ref{cbianchi:transport}) are given by
\begin{subequations}
\begin{eqnarray}
&& \tau \partial_\tau \phi^{(p)}_1 + \frac{1}{2}X_+\phi^{(p)}_2 -\frac{1}{2} X_-\phi_0^{(p)} -p\phi_1^{(p)}= S^{(p)}_1, \label{expanded:cbianchi:transport:1}\\
&& \tau \partial_\tau \phi^{(p)}_2 + \frac{1}{2}X_+\phi^{(p)}_3 -\frac{1}{2} X_-\phi_1^{(p)} -p\phi_2^{(p)}= S^{(p)}_2, \label{expanded:cbianchi:transport:2}\\
&& \tau \partial_\tau \phi^{(p)}_3 +\frac{1}{2} X_+\phi^{(p)}_4 -\frac{1}{2} X_-\phi_0^{(p)} -p\phi_3^{(p)}= S^{(p)}_3, \label{expanded:cbianchi:transport:3}
\end{eqnarray}
\end{subequations}
with $S_j=S_j(u^{(0)},\ldots,u^{(p-1)})$, $j=1,\ldots,3$.

\bigskip
In order to extract detailed information from the above equations one
makes use of an explicit decomposition of the various functions in
terms of the spherical harmonics $\TT{i}{j}{k}$.  We recall the
following lemma which was proved in \cite{Fri98a}.

\begin{lemma}
\label{lemma:propagation:expansion:type}
The following rules for expansion types hold:
\begin{itemize}
\item[(i)] The functions $(c^1_{AB}-\rho x_{AB})^{(p)}$,
  $\upsilon^{(p)}$, $\phi^{(p)}$, $p=1,2,\ldots$ on $\mathcal{I}$
are of expansion type $p-2$, $p-1$, $p$ respectively.

\item[(ii)] The functions $R^{(p)}_i$, $i=0,\ldots,4$ and $S^{(p)}_j$,
  $j=1,2,3$ are of expansion type $p-1$ for $p=1,2,\ldots$.

\item[(iii)] If for a given integer $p\geq 1$ the data for
  $\phi^{(p)}$ on $\mathcal{C}_{a,\kappa}$ are of type $p-1$, then
  $\phi^{(p)}$ on $\mathcal{I}$ is of type $p-1$.

\end{itemize}
\end{lemma}

\subsection{Decomposition in terms of spherical harmonics}

Given the vector $u^{(p)}=(u_1^{(p)},\dots,u^{(p)}_N)$ ---respectively $\upsilon^{(p)}$, $\phi^{(p)}$--- and 
non-negative integers $q$ and $k=0,\dots,2q$ one defines the
\emph{sector} $\mathfrak{S}_{q,k}[u^{(p)}]$ as the
collection of coefficients
\begin{equation*}
%\label{sh:decomposition}
u_{i;2q,k}= (2q+1) \int_{SU(2)} \bar{u}^{(p)}_i \TT{2q}{k}{q-s} \mbox{d}\mu,
\end{equation*} 
where $s$ is the spin-weight of $u^{(p)}_i$, and $\mbox{d}\mu$ is the
\emph{Haar measure} of $SU(2)$. Furthermore, one defines
\[
\mathfrak{S}_{q}[u^{(p)}]= \bigcup_{k=0}^{2q} \mathfrak{S}_{q,k}[u^{(p)}].
\]
A sector will be said to be \emph{vanishing} if $\mathfrak{S}_q[u^{(p)}]=\{ 0 \}$. 

\bigskip
The Weyl spinor of time symmetric, conformally flat initial data is of
expansion type $p-1$ on $\mathcal{C}_{a,\kappa}$. Accordingly, one
writes
\[
\phi_j^{(p)} = \sum_{q=|2-j|}^p \sum_{k=0}^{2q} {\binom{4}{j}}^{-1} a_{j,p;2q,k} \TT{2q}{k}{q-2+j},
\]
with complex ($\tau$-dependent) coefficients $a_{j,p;2q,k}$. The
normalisation factor ${\binom{4}{j}}^{-1}$ has been added for
convenience. The substitution of the latter expression into equations
(\ref{expanded:pbianchi:transport:0})-(\ref{expanded:pbianchi:transport:4})
and
(\ref{expanded:cbianchi:transport:1})-(\ref{expanded:pbianchi:transport:0})
renders equations for the various coefficients $a_{j,p;2q,k}$. In the
cases $p\geq 0$, $q=0$, one finds the equations
\[
a'_{2,p;0,0}=6R_{2,p;0,0},
\]
and 
\[
\tau a'_{2,p;0,0}-p a_{2,p;0,0}=6S_{2,p;0,0}.
\]
If $p\ge 1$, $q=1$, $k=0,\;1,\;2$ one finds
\begin{eqnarray*}
&&  a'_{1,p;2,k} +\frac{1}{3}\beta_2 a_{2,p;2,k} + a_{1,p;2,k}= 4R_{1,p;2,k}, \\
&& a'_{2,p;2,k} +\frac{3}{4}\beta_2 a_{3,p;2,k} -\frac{3}{4}\beta_2 a_{1,p;2,k}= 6R_{2,p;2,k}, \\
&& a'_{3,p;2,k} -\frac{1}{3}\beta_2 a_{2,p;2,k} -a_{3,p;2,k}= 4R_{3,p;2,k},
\end{eqnarray*}
and
\begin{equation*}
\tau a'_{2,p;2,k} +\frac{3}{4}\beta_2 a_{3,p;2,k} + \frac{3}{4}\beta_2 a_{1,p;2,k} -p a_{2,p;2,k}= 6S_{2,p;2,k}.
\end{equation*}
More crucially, one
obtains for $2 \leq p$, $2 \leq q$, $k=0,\ldots,2q$ the equations
\begin{subequations}
\begin{eqnarray}
&& (1+\tau) a'_{0,p;2q,k} + \frac{1}{4}\beta_1a_{1,p;2q,k} -(p-2) a_{0,p;2q,k} = R_{0,p;2q,k}, \label{p0}\\
&& a'_{1,p;2q,k} +\frac{1}{3}\beta_2 a_{2,p;2q,k} -2\beta_1 a_{0,p;2q,k} + a_{1,p;2q,k}= 4R_{1,p;2q,k}, \label{p1}\\
&& a'_{2,p;2q,k} +\frac{3}{4}\beta_2 a_{3,p;2q,k} -\frac{3}{4}\beta_2 a_{1,p;2q,k}= 6R_{2,p;2q,k}, \label{p2}\\
&& a'_{3,p;2q,k} + 2 \beta_1 a_{4,p;2q,k} -\frac{1}{3}\beta_2 a_{2,p;2q,k} -a_{3,p;2q,k}= 4R_{3,p;2q,k}, \label{p3}\\
&& (1-\tau) a'_{4,p;2q,k} -\frac{1}{4}\beta_1 a_{3,p;2q,k} + (p-2) a_{4,p;2q,k}= R_{4,p;2q,k}, \label{p4}
\end{eqnarray}
\end{subequations}
and 
\begin{subequations}
\begin{eqnarray}
&& \tau a'_{1,p;2q,k} +\frac{1}{3}\beta_2 a_{2,p;2q,k} + 2\beta_1 a_{0,p;2q,k} -pa_{1,p;2q,k} = 4S_{1,p;2q,k}, \label{c1}\\
&& \tau a'_{2,p;2q,k} +\frac{3}{4}\beta_2 a_{3,p;2q,k} + \frac{3}{4}\beta_2 a_{1,p;2q,k} -p a_{2,p;2q,k}= 6S_{2,p;2q,k}, \label{c2}\\
&& \tau a'_{3,p;2q,k} + 2\beta_1 a_{4,p;2q,k} +\frac{1}{3}\beta_2 a_{2,p;2q,k} - p a_{3,p;2q,k}= 4S_{3,p;2q,k}, \label{c3}
\end{eqnarray}
\end{subequations}
where 
\[
\beta_1 =\sqrt{(q-1)(q+2)}, \quad \beta_2=\sqrt{q(q+1)},
\]
and $R_{i,p;2q,k}$, $j=0,\ldots,4$ and $S_{j,p;2q,k}$, $i=1,2,3$ are such that 
\begin{equation}
\label{definition:R:S}
R_i^{(p)} = \sum_{q=|2-j|}^p \sum_{k=0}^{2q} R_{j,p;2q,k} \TT{2q}{k}{q-2+j}, \quad S_i^{(p)} = \sum_{q=|2-j|}^p \sum_{k=0}^{2q} S_{j,p;2q,k} \TT{2q}{k}{q-2+j}.
\end{equation}
The functions $R^{(p)}_i$ and $S^{(p)}_j$ contain products of
$\phi^{(p')}$ and $\upsilon^{(p'')}$ for $0\leq p'\leq p-1$,
$0\leq p''\leq p-1$ so that in order to obtain the representation
(\ref{definition:R:S}) one has to linearise products of the form
$\TT{i_1}{j_1}{k_1}\times\TT{i_2}{j_2}{k_2}$ using the formula (\ref{Clebsch-Gordan}).

\bigskip
For latter reference the following result is noted.

\begin{lemma}
\label{lemma:logs:lead:to:logs}
  If the s-jets $J^{(p-1)}_{\mathcal{I}}[\upsilon]$ and
  $J^{(p-1)}_{\mathcal{I}}[\phi]$ have polynomial dependence in $\tau$
  for some $p\geq 1$, then $J^{(p)}_{\mathcal{I}}[\upsilon]$ has also
  polynomial dependence in $\tau$.
\end{lemma}

The proof of this lemma follows directly from the structure of the
transport equation (\ref{upsilon:transport}).

\subsection{Discrete symmetries of the development}
\label{section:discrete:symms}  

It is well known that if a spacetime $(\mathcal{M},g_{\mu\nu})$ is the
development of time symmetric data, then the
spacetime has a discrete time reflexion symmetry: that is one has
$g_{\mu\nu}(t,x^\alpha)=g_{\mu\nu}(-t,x^\alpha)$, where $t$ is a time
function such that $t=0$ yields the slice of time symmetry. However,
in order to discuss the effect of this discrete symmetry on spinorial
objects one has to be more careful as the transformation $t\mapsto -t$
changes the handedness of the canonical orthonormal tetrad associated
to a spin dyad $\{\delta_A\}_{A=0,1}$.

\bigskip In the case of the manifold $\mathcal{M}_{a,\kappa}$, the
discrete transformation $\tau\mapsto -\tau$ induces on the dyad
$\{\delta_A\}_{A=0,1}$ the transitions
\[
\delta_0 \mapsto \delta_1, \quad \delta_1 \mapsto \delta_0,
\]
so that, for example, $\tau_{AA'}\mapsto \tau_{AA'}$. On the other
hand, one has that
\begin{eqnarray*}
&& \epsilon_{AB}\mapsto -\epsilon_{AB}, \\
&& x_{AB}\mapsto x_{AB}, \quad y_{AB}\mapsto -z_{AB}, \quad z_{AB}\mapsto -y_{AB}. \end{eqnarray*}
Similarly, one has that
\begin{eqnarray*}
&& \epsilon^{i}_{ABCD}\mapsto \epsilon^{4-i}_{ABCD}, \quad i=0,\dots,4, \\
&& (\epsilon_{AC}x_{BD}+\epsilon_{BD}x_{AC})\mapsto -(\epsilon_{AC}x_{BD}+\epsilon_{BD}x_{AC}), \\
&& (\epsilon_{AC}y_{BD}+\epsilon_{BD}y_{AC})\mapsto (\epsilon_{AC}z_{BD}+\epsilon_{BD}z_{AC}), \quad (\epsilon_{AC}z_{BD}+\epsilon_{BD}z_{AC})\mapsto (\epsilon_{AC}y_{BD}+\epsilon_{BD}y_{AC}), \\
&& h_{ABCD} \mapsto h_{ABCD}. 
\end{eqnarray*}
From the latter, one deduces the following transition rules on the
components of the Weyl spinor $\phi_{ABCD}$:
\[
\phi_4(\tau)\mapsto \phi_0(-\tau), \quad \phi_3(\tau)\mapsto\phi_1(-\tau), \quad \phi_2(\tau)\mapsto\phi_2(-\tau), \quad \phi_1(\tau)\mapsto \phi_3(-\tau), \quad \phi_0(\tau)\mapsto \phi_4(-\tau).
\]
Furthermore, in order to correct the change of handedness produced by
the transformation $\tau\mapsto -\tau$, one has the following
correspondence rules for the operators $X_\pm$:
\[
X_+\mapsto -X_-, \quad X_-\mapsto -X_+.
\]
In addition, it is noted that
\[
\partial_\tau\phi_4(\tau)\mapsto -\partial_\tau\phi_0(-\tau), \quad \partial_\tau\phi_3(\tau)\mapsto-\partial_\tau\phi_1(-\tau), \quad \partial_\tau\phi_2(\tau)\mapsto-\partial_\tau\phi_2(-\tau).
\]
Combining the above rules together with equations
(\ref{expanded:pbianchi:transport:0})-(\ref{expanded:pbianchi:transport:4})
and
(\ref{expanded:cbianchi:transport:1})-(\ref{expanded:cbianchi:transport:3})
one deduces the following transition rules for their right hand sides:
\begin{eqnarray*}
&& R_0 \mapsto -R_4, \quad R_1 \mapsto -R_3, \quad R_2\mapsto -R_2, \quad R_3\mapsto-R_1, \quad R_4 \mapsto-R_0\\
&& S_1 \mapsto S_3, \quad S_2 \mapsto S_2, \quad S_3\mapsto S_1.
\end{eqnarray*}

\bigskip
For a spacetime with time reflexion symmetry, the aforediscussed transition rules allow to deduce the symmetries satisfied by the various terms appearing in the transport equations (\ref{b0})-(\ref{b4}) and (\ref{bc1})-(\ref{bc3}).

\begin{lemma}
\label{lemma:symms2}
For a spacetime arising from time symmetric initial data, the solutions
to the Bianchi transport equations satisfy the parity conditions
\begin{eqnarray*}
&& \partial^r_\tau a_{4,p;2q,k}(\tau)= (-1)^r \partial^r_\tau a_{0,p;2q,k}(-\tau), \\
&& \partial^r_\tau a_{3,p;2q,k}(\tau)= (-1)^r \partial^r_\tau a_{1,p;2q,k}(-\tau), \\
&& \partial^r_\tau a_{2,p;2q,k}(\tau)= (-1)^r \partial^r_\tau a_{2,p;2q,k}(-\tau),
\end{eqnarray*}
and also
\begin{eqnarray*}
&& \partial^r_\tau R_{4,p;2q,k}(\tau)= (-1)^{r+1}\partial^r_\tau R_{0,p;2q,k}(-\tau), \\
&& \partial^r_\tau R_{3,p;2q,k}(\tau)= (-1)^{r+1}\partial^r_\tau R_{1,p;2q,k}(-\tau), \\
&& \partial^r_\tau R_{2,p;2q,k}(\tau)= (-1)^{r+1}\partial^r_\tau R_{2,p;2q,k}(-\tau), \\
&& \partial^r_\tau S_{3,p;2q,k}(\tau)= (-1)^r \partial^r_\tau S_{1,p;2q,k}(-\tau), \\
&& \partial^r_\tau S_{2,p;2q,k}(\tau)= (-1)^r \partial^r_\tau S_{2,p;2q,k}(-\tau).
\end{eqnarray*}
\end{lemma}

\subsection{A procedure to solve the Bianchi transport equations}
\label{section:alternative:reduced}

A first analysis of the structure of the solutions to equations
(\ref{p0})-(\ref{p4}) has been given in \cite{Fri98a}. In particular,
in the aforementioned reference a procedure was given by means of
which the constraint equations (\ref{c1})-(\ref{c3}) are used to
eliminate the unknowns $a_{1,p;2q,k}$, $a_{2,p;2q,k}$, $a_{3,p;2q,k}$
so that to find a solution to the transport equations
(\ref{p0})-(\ref{p4}) it is only necessary to solve a \emph{reduced
system} involving $a_{0,p;2q,k}$ and $a_{4,p;2q,k}$. The remaining
coefficients are then obtained by means of purely algebraic
manipulations.

\medskip
For the purpose of the present investigation it will turn out to be
more convenient to consider an alternative procedure to find
solutions of the equations (\ref{p0})-(\ref{p4}). Again, equations
(\ref{c1})-(\ref{c3}) will be used to obtain a reduced system. But in
this case, the system will involve $a_{1,p;2q,k}$ and
$a_{3,p;2q,k}$. Due to the formal similarity with the reduced systems
obtained in \cite{Val07b}, this type of reduced system will be called
\emph{Maxwell-like}. The reasons to prefer this approach over the one
put forward in \cite{Fri98a} will be explained towards the end of this
section.

\medskip
In the following discussion, the values of the
indices $p$, $q$ and $k$ are considered as fixed. In order to ease the
formulae, obvious indices will be suppressed in the following.

\medskip
One can construct a Maxwell-like propagation system involving the
coefficients $a_1$, $a_2$ and $a_3$ by considering the sum of
equations (\ref{p1}) and (\ref{c1}), equation (\ref{p2}) and the
difference of equations (\ref{p3}) and (\ref{c3}). The resulting equations are
given by
\begin{subequations}
\begin{eqnarray}
&& (1+\tau) a'_1 +\frac{2}{3}\beta_2 a_2 -(p-1) a_1 = 2R_1 + 4S_1, \label{mp1}\\
&& a'_2 +\frac{3}{4}\beta_2 a_3 - \frac{3}{4}\beta_2 a_1 = 6 R_2, \label{mp2}\\
&& (1-\tau) a'_3 -\frac{2}{3}\beta_2 a_2 +(p-1) a_3 = 2R_3 -4S_3, \label{mp3}
\end{eqnarray}
\end{subequations}
where the prime denotes differentiation with respect to $\tau$. 
Note that the above equations do not contain $a_0$ or $a_4$. The
associated Maxwell-like constraint equation is given simply by
equation (\ref{c2}). Namely
\begin{equation}
\tau a'_2 +\frac{3}{4}\beta_2 a_3 + \frac{3}{4}\beta_2 a_1 -pa_2 =6S_2. \label{mc}
\end{equation}
From equations (\ref{mc}) and (\ref{mp2}) one obtains the algebraic relation
\begin{equation}
\frac{3}{4}\beta_2(1-\tau) a_3 + \frac{3}{4}\beta_2(1+\tau) a_1 -pa_2 = 6S_2 -6\tau R_2, \label{algebraic}
\end{equation} 
which can be used, in turn, to eliminate $a_2$ from both (\ref{mp1}) and (\ref{mp3}) so that
\begin{subequations}
\begin{eqnarray}
&& (1+\tau) a'_1 + \left(1-p + \frac{1}{2p}q(q+1)(1+\tau) \right)a_1 + \frac{1}{2p}q(q+1)(1-\tau) a_3 \nonumber \\
&& \hspace{7cm} = 2R_1 + 4S_1 + \frac{4}{p}\beta_2 S_2 - \frac{4}{p}\beta_2 \tau R_2, \label{reduced:mp1} \\
&& (1-\tau) a'_3 -\frac{1}{2p} q(q+1)(1+\tau) a_1 + \left(p-1 -\frac{1}{2p}q(q+1)(1-\tau)  \right) a_3  \nonumber \\
&& \hspace{7cm} = 2R_3 -4S_3 -\frac{4}{p}\beta_2 S_2 +\frac{4}{p}\beta_2 \tau R_2. \label{reduced:mp3}
\end{eqnarray}
\end{subequations}
Equations (\ref{reduced:mp1})-(\ref{reduced:mp3}) will be referred
to as the \emph{reduced Maxwell-like system for}
$\mathfrak{S}_{q,k}[\phi^{(p)}]$. Given a solution $a_1$, $a_3$ to
the reduced system (\ref{reduced:mp1})-(\ref{reduced:mp3}), the
coefficient $a_2$ is obtained from equation (\ref{algebraic}) by means
of an algebraic manipulation, while $a_0$ and $a_4$ are obtained
as a solution of the ordinary differential equations (\ref{p0}) and
(\ref{p4}).

\bigskip
In order to ease the subsequent discussion, the system (\ref{reduced:mp1}) and (\ref{reduced:mp3}) is written in matricial form as
\begin{equation}
y'(\tau) =A(\tau) y(\tau) + b(\tau), \label{reduced:system}
\end{equation}
with
\[
A(\tau) \equiv \left(
\begin{array}{cc}
 \displaystyle -\frac{1}{1+\tau}\left(1-p + \frac{1}{2p}q(q+1)(1+\tau) \right) & \displaystyle  -\frac{1}{2p}q(q+1)\frac{1-\tau}{1+\tau} \\
\displaystyle \frac{1}{2p}q(q+1) \frac{1+\tau}{1-\tau} & \displaystyle -\frac{1}{1-\tau}\left(p-1 -\frac{1}{2p}q(q+1)(1-\tau)  \right)
\end{array}
\right),
\]
and
\[
y(\tau) \equiv
\left(
\begin{array}{c}
a_1(\tau) \\
a_3(\tau)
\end{array}
\right),
\quad 
b(\tau) \equiv
\left(
\begin{array}{c}
\displaystyle \frac{1}{1+\tau}F_1(\tau) \\
\displaystyle \frac{1}{1-\tau}F_3(\tau)
\end{array}
\right),
\]
with
\begin{eqnarray*}
&&F_1 \equiv 2R_1 + 4S_1 + \frac{4}{p}\beta_2 S_2 - \frac{4}{p}\beta_2 \tau R_2, \\
&&F_3 \equiv 2R_3 -4S_3 -\frac{4}{p}\beta_2 S_2 +\frac{4}{p}\beta_2 \tau R_2. 
\end{eqnarray*}
It follows from the discussion of section (\ref{section:discrete:symms}) that
\[
F_3(\tau) =-F_1(-\tau) \equiv -F^s_{1}(\tau). 
\]

\bigskip
Following the ideas of the discussion in \cite{Val07b}, it is possible to
find a fundamental matrix for the system (\ref{reduced:system}). One
obtains:
\[
X_{p,q} \equiv
\left(
\begin{array}{cc}
Q_1 & (-1)^{q+1} Q_3 \\
(-1)^{q+1} Q^s_{3} &  Q^s_{1}
\end{array}
\right),
\]
with
\begin{subequations}
\begin{eqnarray}
&& Q_1(\tau) \equiv \left(\frac{1-\tau}{2} \right)^{p+1} P_{q-1}^{(p+1,1-p)}(\tau), \label{Q1}\\
&& Q_3(\tau) \equiv \left(\frac{1+\tau}{2} \right)^{p-1} P_{q+1}^{(-p-1,p-1)}(\tau), \label{Q3}
\end{eqnarray}
\end{subequations}
where $P^{(\alpha,\beta)}_n(\tau)$, $n$ a non-negative integer,
denotes a \emph{Jacobi polynomial} ---see e.g. \cite{Sze78} for
definitions and properties. Furthermore:
\[
Q^s_{1}(\tau) \equiv Q_1(-\tau), \quad Q^s_{3}(\tau) \equiv Q_{3}(-\tau).
\]
The determinant of the fundamental matrix $X_{p,q}$ (the Wronskian) is given by
\begin{equation}
\label{Wronskian}
\det X_{p,q} = W_0 (1-\tau^2)^{p-1},
\end{equation}
with $W_0$ a constant. Consequently, the inverse $X_{p,q}^{-1}$ is given by
\[
X^{-1}_{p,q} = \frac{1}{W_0}
\left(
\begin{array}{cc}
\displaystyle \frac{(1+\tau)^2}{(1-\tau)^{p-1}} K_{p,q} & \displaystyle \frac{(-1)^q}{(1-\tau)^{p-1}}L_{p,q} \\
\displaystyle \frac{(-1)^q}{(1+\tau)^{p-1}} M_{p,q} & \displaystyle \frac{(1-\tau)^2}{(1+\tau)^{p-1}}N_{p,q},
\end{array}
\right)
\]
where $K$, $L$, $M$ and $N$ are shorthand for the following Jacobi polynomials
\begin{eqnarray*}
&& K_{p,q}(\tau) \equiv  \frac{1}{2^{p+1}} P_{q-1}^{(p+1,1-p)}(-\tau), \\
&& L_{p,q}(\tau) \equiv  \frac{1}{2^{p-1}} P_{q+1}^{(-p-1,p-1)}(\tau), \\
&& M_{p,q}(\tau) \equiv  \frac{1}{2^{p-1}}P_{q+1}^{(-p-1,p-1)}(-\tau), \\
&& N_{p,q}(\tau) \equiv  \frac{1}{2^{p+1} }P_{q-1}^{(p+1,1-p)}(\tau).
\end{eqnarray*}

\bigskip
The solution to system (\ref{reduced:system}) is given by
\begin{equation}
\label{reduced:system:matrix}
y(\tau) = X(\tau) X^{-1}(0) y(0) + X(\tau) \int_0^\tau X^{-1}(s) b(s) \mbox{d}s.
\end{equation}

One can verify that the procedure described in the previous lines does
indeed provide a solution to equations (\ref{p0})-(\ref{p4}) and
(\ref{c1})-(\ref{c3}). The argument is a follows:
\begin{enumerate}
\item[(i)] One solves the
Maxwell-like reduced system (\ref{reduced:mp1})-(\ref{reduced:mp3})
using formula (\ref{reduced:system:matrix}) or any other method.

\item[(ii)] Next, one substitutes the values of the coefficients $a_1$ and $a_3$ obtained in this way into the evolution (\ref{mp2}). This equation can be
solved for $a_2$ by a direct integration. 

\item[(iii)] Given the evolution equations (\ref{reduced:mp1}),
(\ref{mp2}) and (\ref{reduced:mp3}) one can produce an argument to
show the propagation of the constraint equation (\ref{mc}) ---if (\ref{mc}) is satisfied initially for $\tau=0$, then it is
satisfied at later times.

\item[(iv)] Equations (\ref{mp2}) and (\ref{mc}) imply the algebraic
condition (\ref{algebraic}). The latter, together with
(\ref{reduced:mp1}) and (\ref{reduced:mp3}) imply the evolution
equations (\ref{mp1}) and (\ref{mp3}).

\item[(v)] One substitutes the coefficients $a_1$, $a_2$ and $a_3$ into the
Bianchi propagation equations (\ref{p0}) and (\ref{p4}). Again, these
equations can be solved for the coefficients $a_0$ and $a_4$ by means
of a direct integration. 

\item[(vi)] It can be shown that the evolution equations
(\ref{p0})-(\ref{p4}) imply the propagation of the
constraint equations (\ref{c1}) and (\ref{c3}).
\end{enumerate}

\subsection{General properties of the solutions to the reduced Maxwell-like system}
\label{section:general:properties:maxwell}

Let $C^\omega(a,b)$ denote the set of analytic functions on the
interval $(a,b)$. From formula (\ref{reduced:system:matrix}) one
obtains the following result.

\begin{proposition}
\label{proposition:logarithmic:solutions}
If the components of the vector $b(\tau)$ are polynomials, then the
solutions $a_1$, $a_3$ to the Maxwell-like reduced system
(\ref{reduced:system}) will be either polynomial in $\tau$ or of the
form 
\begin{eqnarray*}
&& a_1= P(\tau) + \frac{1}{W_0} Q_1(\tau) \ln(1-\tau) + \frac{1}{W_0}(-1)^{q+1} Q_3 \ln(1+\tau), \\
&& a_3= P^s(\tau) + \frac{1}{W_0} (-1)^{q+1} Q^s_3(\tau) \ln(1-\tau)+ \frac{1}{W_0} Q^s_1(\tau) \ln(1+\tau).
\end{eqnarray*}
where $P(\tau)$ is a polynomial in $\tau$ and $Q_1(\tau)$ and
$Q_3(\tau)$ are as in (\ref{Q1}) and (\ref{Q3}). These last
solutions are of class $C^\omega(-1,1) \cap C^{p-1}[-1,1]$.
\end{proposition}

\textbf{Proof.} The crucial observation is to note that the particular
form of the Wronskian (\ref{Wronskian}) implies that the partial
fraction decomposition of the entries in the integrand,
$X^{-1}_{p,q}b$, of formula (\ref{reduced:system:matrix}) contains
negative integer powers of $(1\pm\tau)$. The terms $(1\pm \tau)^{-1}$
will integrate to $\ln|1\pm\tau|$. The terms with $(1\pm \tau)^{-r}$,
$r\geq 2$ integrate to terms of the same form. Multiplication by the
matrix $X_{p,q}$ removes these rational terms. \hfill $\Box$

\bigskip
\textbf{Remark 1.} In order to have polynomial-only solutions,
some special cancellations should occur in the partial
fraction decomposition of the entries of $X^{-1}_{p,q}b$. In the
sequel it will be shown that these cancellations do occur for particular combinations of the mutiindices $(p;q,k)$. 

\bigskip
\textbf{Remark 2.} The particular form of the Wronskian
(\ref{Wronskian}) is the reason why the approach described in this
section has been preferred to the one originally described in
\cite{Fri98a}. The Wronskian of the reduced system advocated in the
aforementioned reference is of the form
\[
c f(\tau) (1-\tau)^{p-2},
\]
with $c$ a constant and 
\[
f(\tau)\equiv 2(p+1)(p-1)-(q-1)(q+2)(1-\tau^2).
\]
The partial fraction decomposition of the entries of the corresponding matrix product $X^{-1}_{p,q}b$ will contain terms of the form
\[
\frac{\alpha \tau + \beta}{ 2(p+1)(p-1)-(q-1)(q+2)(1-\tau^2)},
\]
for $\alpha$, $\beta$ some constants. These will integrate to give
multiples of terms of the form
\[
\ln\left|2(p+1)(p-1)-(q-1)(q+2)(1-\tau^2)\right|, \quad \arctan\left(\sqrt{\frac{(q-1)(q+2)}{2(p+1)(p-1)-(q-1)(q+2)}} \tau \right).
\]
As a result of proposition \ref{proposition:logarithmic:solutions},
one is expecting solutions consisting only of polynomials and
$\ln|1\pm\tau|$. Thus, there are some non-trivial cancellations in
formula (\ref{reduced:system:matrix}) that need to be explained by
means of some further arguments.

\subsection{The Case $p=q$ and the regularity condition at $i$}
\label{section:coton}
As first pointed out in \cite{Fri98a}, for $p\geq 2$ if $q=p$, then the vector $b$ appearing in formula (\ref{reduced:system:matrix}) is such that
$\mathfrak{S}_{p,k}[b]={0}$, 
so that the solution to the reduced system (\ref{reduced:system}) is
given entirely by the solution to the homogeneous problem. From here,
it is possible to identify conditions on the initial data so that the
solutions to the Bianchi transport equations for these particular
sectors extend smoothly through the critical sets $\mathcal{I}^\pm$.
In particular, one has the following result ---theorem 8.2 in \cite{Fri98a}.

\begin{theorem}
  Given vacuum initial data which is time symmetric and analytic in a
  neighbourhood of infinity, the solution to the regular finite
  initial value problem is smooth through $\mathcal{I}^\pm$ only if
  the condition
\begin{equation}
\label{Helmut:regularity:condition}
D_{(E_pF_p}\cdots D_{E_1F_1} b_{ABCD)}(i)=0, \quad p=1,2,\ldots.
\end{equation}
If this condition is violated at some order $p'$, then the solution
will develop logarithmic singularities in
$\mathfrak{S}_{p'}[\phi^{(p')}]$ at $\mathcal{I}^\pm$.
\end{theorem}

In the previous result $b_{ABCD}$ denotes the spinorial counterpart of the Cotton tensor. The Cotton tensor is given by
\[
k_{pij} \equiv D_j l_{ip}-D_i l_{jp}, \quad  l_{ij}\equiv s_{ij} + \frac{1}{12} rh_{ij},
\]
where $s_{ij}$ denotes the tracefree part of the Ricci tensor of the
initial 3-metric $h_{ij}$. Associated to $k_{pij}$ one has the spinor
$k_{ABCDEF}$, which because of the symmetries of the tensorial
counterpart can be written as
\[
k_{ABCDEF} = b_{ABCE} \epsilon_{DF} + b_{ABDF} \epsilon_{CE}.
\]

\medskip
\textbf{Remark.} Time symmetric data which is conformally flat in a
neighbourhood of infinity satisfies the condition
(\ref{Helmut:regularity:condition}) trivially to all orders.

\section{The solutions to the transport equations for data which is Schwarzschildean up to a certain order}
\label{section:punchline}

In this section the main analysis of the present work is
presented. The key idea behind is to analyse the solutions to the
transport equation at spatial infinity for initial data sets which are
Schwarzschildean up to a certain order.

\subsection{Summary of the results of  reference \cite{Val04a}}

In order to motivate the analysis of this section, a brief summary of
the results of \cite{Val04a} is presented. In that reference one
assumes a function $W$ of the form
\[
W= \frac{m}{2} + \sum_{p=2}^\infty \sum_{k=0}^{2p} \frac{1}{p!} w_{p;2p,k} \TT{2p}{k}{p} \rho^p.
\]
In fact, the form of $W$ assumed in \cite{Val04a} is slightly more
general than this as it includes terms $w_{1,2,k}$, $k=0,1,2$, which
---as seen in section \ref{section:conformally:flat:data}--- can
always be removed by choosing the centre of mass properly. Using
scripts in the computer algebra system {\tt Maple V} one can calculate
the explicit solutions to the transport equations
(\ref{upsilon:transport})-(\ref{cbianchi:transport}) via the
decomposition in terms of spherical harmonics discussed in section
\ref{section:further:properties}. The solutions for the orders
$p=0,1,2,3$ have been calculated in \cite{Fri98a,FriKan00}. These
solutions have polynomial dependence in $\tau$, and thus, extend
smoothly through the critical sets, although as seen in proposition
\ref{proposition:logarithmic:solutions}, the solutions could have had
logarithmic singularities. The use of computer algebra methods allows to go
beyond this point and to calculate further orders in the expansions. For $p=4$
one finds again that the solutions have polynomial dependence on
$\tau$. However, for $p=5$ the situation is different. One finds that
the sectors $\mathfrak{S}_2[\phi^{(5)}]$ has solutions with logarithmic
terms of the form given by proposition
\ref{proposition:logarithmic:solutions}. An important observation is
that these logarithmic solutions do not appear if one chooses a
function $W$ for which
\[
w_{2,4,k}=0, \quad k=0,\ldots, 4.
\]
In the terminology of section \ref{section:conformally:flat:data} this
means that the data is Schwarzschildean up to order $p=2$. Assuming
that this is the case, one can proceed further with the expansions. At order
$p=6$ one finds logarithmic solutions only in the sectors
$\mathfrak{S}_3[\phi^{(6)}]$. The logarithmic solutions can be avoided
by considering data such that
\[
w_{3,6,k}=0, \quad k=0,\ldots, 6,
\]
that is, data which is Schwarzschildean up to order $p=3$. From these
results one can already infer a pattern which has been confirmed to all
the orders for which the calculations have been carried out. The
calculations reported in \cite{Val04a} are carried out up to order
$p=9$, but there is no reason why ---besides computing power--- the
calculations cannot be carried out any further.

\bigskip
The pattern inferred from the analysis in \cite{Val04a} is as
follows. Assume that one has an initial data set which is
Schwarzschildean up to order $p=\pb$. Then, the solutions to the
transport equations for $p\leq \pb$ will only contain the sectors
$\mathfrak{S}_0[u^{(p)}]$. These will coincide with the sectors implied
by the s-jet $J^{(\pb)}_{\mathcal{I}}[u^\pb]$ of the Schwarzschild
spacetime. At order $p=\pb+1$ the non-vanishing sectors are
\[
\mathfrak{S}_0[u^{(\pb+1)}], \quad \mathfrak{S}_{\pb+1}[u^{(\pb+1)}].
\]
 Both of them are polynomial in
$\tau$. At order $p=\pb+2$ the non-vanishing sectors are
\[
\mathfrak{S}_0[u^{(\pb+2)}], \quad \mathfrak{S}_{\pb+1}[u^{(\pb+2)}], \quad 
\mathfrak{S}_{\pb+2}[u^{(\pb+2)}].
\] 
Again, all the non-vanishing
sectors are polynomial in $\tau$.  At order $p=\pb+3$ the non-vanishing
sectors are 
\[
\mathfrak{S}_0[u^{(\pb+3)}], \quad  \mathfrak{S}_{\pb+1}[u^{(\pb+3)}],\quad  \mathfrak{S}_{\pb+2}[u^{(\pb+3)}], \quad \mathfrak{S}_{\pb+3}[u^{(\pb+3)}].
\]
All the non-vanishing
sectors are polynomial in $\tau$. Finally, at order $p=\pb+4$ one has
the following non-vanishing sectors: 
\[
\mathfrak{S}_0[u^{(\pb+4)}], \quad \mathfrak{S}_{\pb+1}[u^{(\pb+4)}], \quad 
\mathfrak{S}_{\pb+2}[u^{(\pb+4)}], \quad  \mathfrak{S}_{\pb+3}[u^{(\pb+4)}], \quad
\mathfrak{S}_{\pb+4}[u^{(\pb+4)}].
\]
 The sectors
$\mathfrak{S}_0[u^{(\pb+4)}]$, $\mathfrak{S}_{\pb+2}[u^{(\pb+4)}]$,
$\mathfrak{S}_{\pb+3}[u^{(\pb+4)}]$ and
$\mathfrak{S}_{\pb+4}[u^{(\pb+4)}]$ have polynomial dependence in
$\tau$, while $\mathfrak{S}_{\pb+1}[u^{(\pb+4)}]$ will have logarithmic
singularities of the type indicated in proposition
\ref{proposition:logarithmic:solutions}. Furthermore, the solutions
will be logarithmic free if
\[
w_{\pb+1,2(\pb+1),k} =0, \quad k=0,\ldots, 2(\pb+1). 
\]
This pattern will be effectively proved in the sequel. It readily
suggests an inductive procedure to prove the main theorem presented in
the introductory section. The calculations in \cite{Val04a} are the
base step of this inductive procedure.

\subsection{Properties of data which is Schwarzschildean up to order $p=p_\bullet$}
We start with some generic observations which will be used
systematically in the sequel. Assume that the function $W$, appearing in 
expression (\ref{theta_parametrised}) for the conformal factor
$\vartheta$, is of the form
\begin{equation}
\label{W:quasi:Schw}
W= \frac{m}{2} + \sum_{p=\pb+1}^\infty \sum_{k=0}^{2p} \frac{1}{p!} w_{p;2p,k} \TT{2p}{k}{p} \rho^p,
\end{equation}
so that the initial data is Schwarzschildean up to order $\pb$. 
The function $W$ ---and hence also the coefficients
$w_{p;2p,k}$, $\pb \leq p$, $k=0,\ldots,2p$--- appears non-linearly in
the expression
\[
\Omega= \frac{\rho^2}{(1+\rho W)^2}
\]
 for the conformal factor $\Omega$ and, moreover, in the
expressions for the initial data for the spinors $\phi_{ABCD}$ and
$\Theta_{ABCD}$ on $\mathcal{C}_{a,\kappa}$ ---see equations
(\ref{data1}) and (\ref{data2}). Hence, when calculating the
normal expansions of $\phi_{ABCD}$ and
$\Theta_{ABCD}$ one
encounters products of the form $\TT{i_1}{j_1}{k_1} \times \TT{i_2}{j_2}{k_2}$
with $\TT{i_1}{j_1}{k_1},\; \TT{i_2}{j_2}{k_2}\neq \TT{0}{0}{0}$ which
have to be linearised ---that is, expressed as a linear combination of
other functions $\TT{i}{j}{k}$. This linearisation procedure is
extremely cumbersome and involves the use of the Clebsch-Gordan
coefficients of $SU(2,\Complex)$. Remarkably, it turns out that if one
only considers expansions up to order $p=\pb+4$ ---which is what will
be required in the present analysis--- these higher order products do
not arise. An inspection renders the following result.

\begin{lemma}
\label{quasi:Schw:data}
For initial data sets which are Schwarzschildean up to order $p=\pb$
in $\mathcal{B}_a$ one has that on $\mathcal{C}_{a,\kappa}$
\[
\phi_{ABCD}-\phi_{ABCD}^\bullet=\O(\rho^{\pb+1}), \quad \Theta_{ABCD}-\Theta^\bullet_{ABCD}=\O(\rho^{\pb+2}),
\]
where $\phi_{ABCD}^\bullet$ and $\Theta^\bullet_{ABCD}$ denote,
respectively, the Weyl and Ricci spinors of the Schwarzschild
data. Moreover, if $\pb>3$, then
\begin{eqnarray*}
&& \phi_i= \phi_i^\bullet + \frac{1}{(\pb+1)!}\sum_{k=0}^{2\pb+2} \tilde{w}_{\pb+1;2\pb+2,k}\TT{2\pb+2}{k}{\pb+1-i}\rho^{\pb+1} \\
&& \hspace{1.5cm} + \frac{1}{(\pb+2)!}\left(\sum_{k=0}^{2\pb+2} \tilde{w}_{\pb+1;2\pb+2,k}\TT{2\pb+2}{k}{\pb+3-i} + \sum_{k=0}^{2\pb+4} \tilde{w}_{\pb+2;2\pb+4,k}\TT{2\pb+4}{k}{\pb+4-i} \right)\rho^{\pb+2} \\
&& \hspace{1.5cm} + \frac{1}{(\pb+3)!}\left(\sum_{k=0}^{2\pb+4} \tilde{w}_{\pb+2;2\pb+4,k}\TT{2\pb+4}{k}{\pb+4-i} + \sum_{k=0}^{2\pb+6} \tilde{w}_{\pb+3;2\pb+6,k}\TT{2\pb+6}{k}{\pb+5-i} \right)\rho^{\pb+3} \\
&& \hspace{1.5cm} + \frac{1}{(\pb+4)!}\left(\sum_{k=0}^{2\pb+6} \tilde{w}_{\pb+3;2\pb+6,k}\TT{2\pb+6}{k}{\pb+5-i} + \sum_{k=0}^{2\pb+8} \tilde{w}_{\pb+4;2\pb+8,k}\TT{2\pb+8}{k}{\pb+6-i} \right)\rho^{\pb+4} \\
&& \hspace{1.5cm}+\O(\rho^{\pb+5}),
\end{eqnarray*}
where 
\[
\tilde{w}_{\pb+i;2\pb+2i,k}= c_{\pb+i;2\pb+2i,k}w_{\pb+i;2\pb+2i,k},
\]
with $c_{\pb+i;2\pb+2i,k}$ some numerical coefficients which can be explicitly calculated, and
\begin{eqnarray*}
&&\phi^\bullet_i= -6m, \quad i=2, \\
&&\phi^\bullet_i=0, \quad i\neq 2.
\end{eqnarray*}
A similar expansion holds for the components of $\Theta_{ABCD}$.
\end{lemma}

\bigskip
The following result, which can be proved by direct inspection, shows
that the calculations described in the present work do not require the
calculation of complicated $SU(2,\Complex)$ Clebsch-Gordan
coefficients.

\begin{lemma}
\label{lemma:noClebsch-Gordan}
For the class of initial data under consideration, the terms
\begin{eqnarray*}
&& \sum_{j=1}^{p-1}\binom{p}{j}\left(Q(v^{(j)},v^{(p-j)})+ L^{(j)}\phi^{(p-j)}\right), \\
&& \sum_{j=1}^p
\binom{p}{j}\left(B(\Gamma_{ABCD}^{(j)})\phi^{(p-j)}-A^{AB}(c^\mu_{AB})^{(j)}\partial_\mu
 \phi^{(p-j)}\right), \\
&& \sum_{j=1}^p\binom{p}{j}\left(H(\Gamma^{(j)}_{ABCD})\phi^{(p-j)}-F^{AB}(c^\mu_{AB})^{(j)}\partial_\mu \phi^{(p-j)}\right),  
\end{eqnarray*}
with $p=\pb+1,\ldots,\pb+4$ and $\pb\geq 3$, appearing, respectively,
in the transport equations (\ref{upsilon:transport}),
(\ref{pbianchi:transport}) and (\ref{cbianchi:transport}) contain,
before linearisation only products of the form $\TT{0}{0}{0}\times
\TT{i}{j}{k}$.
\end{lemma}

\bigskip
A direct observation is the following.

\begin{lemma}
  Let $J^{(p)}_{\mathcal{I}}[\upsilon]$ and
  $J^{(p)}_{\mathcal{I}}[\phi]$ be the s-jets of order $p$ arising from
  initial data with ADM mass $m$ which is Schwarzschildean up to order
  $p=\pb$. Then one has that
\[
 J^{(p)}_\mathcal{I}[\upsilon]=J^{(p)}_\mathcal{I}[\upsilon_{\bullet}], \quad J^{(p)}_\mathcal{I}[\phi]=J^{(p)}_\mathcal{I}[\phi_{\bullet}]
,
 \]
for $p=0,\ldots,\pb$, where $J^{(p)}_\mathcal{I}[\upsilon_{\bullet}]$ and
$J^{(p)}_\mathcal{I}[\phi_{\bullet}]$ are
the s-jets of order $p$ arising from exactly Schwarzschildean data with
ADM mass $m$. The first difference between these two sets of jets
arises at order $p=\pb+1$.
\end{lemma}

\noindent
\textbf{Remark.} An inspection of the terms discussed in lemma
\ref{lemma:noClebsch-Gordan} for $p=\pb+4$ shows that the present
discussion requires at most the explicit knowledge of the s-jet
$J^{(4)}[u_{\bullet}]$ of the Schwarzschild spacetime.

\bigskip
In order to appreciate the following arguments, the
non-vanishing sectors in the s-jets
$J^{(\pb+5)}_\mathcal{I}[\upsilon-\upsilon_\bullet]$ and
$J^{(\pb+5)}_\mathcal{I}[\phi-\phi_\bullet]$ will be listed. This list
can be deduced from lemma \ref{lemma:propagation:expansion:type}.

\begin{itemize}
\item At order $p=\pb+1$:
\begin{eqnarray*}
&& \mathfrak{S}_0[\upsilon^{(\pb+1)}], \\
&& \mathfrak{S}_0[\phi^{(\pb+1)}], \; \mathfrak{S}_{\pb+1}[\phi^{(\pb+1)}].
\end{eqnarray*}

\item At order $p=\pb+2$:
\begin{eqnarray*}
&& \mathfrak{S}_0[\upsilon^{(\pb+2)}], \; \mathfrak{S}_{\pb+1}[\upsilon^{(\pb+2)}], \\
&& \mathfrak{S}_0[\phi^{(\pb+2)}], \; \mathfrak{S}_{\pb+1}[\phi^{(\pb+2)}], \; \mathfrak{S}_{\pb+2}[\phi^{(\pb+2)}]. 
\end{eqnarray*}

\item At order $p=\pb+3$:
\begin{eqnarray*}
&& \mathfrak{S}_0[\upsilon^{(\pb+3)}], \; \mathfrak{S}_{\pb+1}[\upsilon^{(\pb+3)}], \; \mathfrak{S}_{\pb+2}[\upsilon^{(\pb+3)}], \\
&& \mathfrak{S}_0[\phi^{(\pb+3)}], \; \mathfrak{S}_{\pb+1}[\phi^{(\pb+3)}], \; \mathfrak{S}_{\pb+2}[\phi^{(\pb+3)}], \; \mathfrak{S}_{\pb+3}[\phi^{(\pb+3)}].
\end{eqnarray*}

\item At order $p=\pb+4$:
\begin{eqnarray*}
&& \mathfrak{S}_0[\upsilon^{(\pb+4)}], \; \mathfrak{S}_{\pb+1}[\upsilon^{(\pb+4)}], \; \mathfrak{S}_{\pb+2}[\upsilon^{(\pb+4)}], \; \mathfrak{S}_{\pb+3}[\upsilon^{(\pb+4)}]  \\
&& \mathfrak{S}_0[ \phi^{(\pb+4)}], \; \mathfrak{S}_{\pb+1}[ \phi^{(\pb+4)}], \; \mathfrak{S}_{\pb+2}[ \phi^{(\pb+4)}], \; \mathfrak{S}_{\pb+3}[ \phi^{(\pb+4)}], \; \mathfrak{S}_{\pb+4}[ \phi^{(\pb+4)}].
\end{eqnarray*}

\end{itemize}   

\bigskip
\noindent
\textbf{Remark.} From lemma (\ref{lemma:noClebsch-Gordan}) it follows that
there is no mixing between the various sectors arising at each order.
Thus, it is only necessary to carry out a discussion of the solutions
in the difference jet $J^{(\pb+5)}_\mathcal{I}[u-u_\bullet]$ for the sectors
$\mathfrak{S}_{\pb+1}$.  The analysis of the corresponding sectors 
$\mathfrak{S}_{\pb+2}$, $\mathfrak{S}_{\pb+3}$, $\mathfrak{S}_{\pb+4}$
and $\mathfrak{S}_{\pb+5}$ can, in principle, be obtained from that of
the sector $\mathfrak{S}_{\pb+1}$ by performing, respectively, the
formal replacements $\pb\mapsto \pb+1$, $\pb\mapsto \pb+2$,
$\pb\mapsto\pb+3$ and $\pb\mapsto \pb+4$.

\bigskip
\noindent
\textbf{Warning.} In order to improve the readability, obvious strings
of subindices will be omitted in the sequel.

\subsection{Properties of the transport equations for $p=\pb+1$}
\label{section:pp1}

The solutions to the transport equations at order $p=\pb+1$ can be
essentially read from the original analysis carried out in
\cite{Fri98a}. Using lemma (\ref{lemma:propagation:expansion:type}) it
follows that $\upsilon^{(\pb+1)}$ contains no contribution to the
sectors $\mathfrak{S}_{\pb+1}$. The only non-vanishing sector in
$\upsilon^{(\pb+1)}$ is $\mathfrak{S}_0$, which coincides with the
Schwarzschildean solution. The (non-vanishing) contribution of
$\phi^{(\pb+1)}$ to the sectors $\mathfrak{S}_{\pb+1}$ is given by:
\begin{subequations}
\begin{eqnarray}
&& \hspace{-1cm} a_{0,\pb+1;2(\pb+1),k}= A_{0,k}(1-\tau)^{\pb+3}(1+\tau)^{\pb-1}, \label{Weyl:pp1_0}\\
&& \hspace{-1cm} a_{1,\pb+1;2(\pb+1),k}= A_{1,k} (1-\tau)^{\pb+2}(1+\tau)^{\pb}, \label{Weyl:pp1_1} \\
&& \hspace{-1cm} a_{2,\pb+1;2(\pb+1),k}= A_{2,k} (1-\tau)^{\pb+1}(1+\tau)^{\pb+1}, \\
&& \hspace{-1cm} a_{3,\pb+1;2(\pb+1),k}= A_{1,k} (1-\tau)^{\pb} (1+\tau)^{\pb+2}, \label{Weyl:pp1_3}\\
&& \hspace{-1cm} a_{4,\pb+1;2(\pb+1),k}= A_{0,k} (1-\tau)^{\pb-1}(1+\tau)^{\pb+3}, \label{Weyl:pp1_4}
\end{eqnarray}
\end{subequations}
with $k=0,\ldots,2(\pb+1)$ and
\begin{eqnarray*}
&& A_{0,k} \equiv -\sqrt{\pb(\pb+1)(\pb+2)(\pb+3)}w_{\pb+1;2(\pb+1),k}, \\
&& A_{1,k} \equiv -4 (\pb+3)\sqrt{(\pb+1)(\pb+2)}w_{\pb+1;2(\pb+1),k}, \\
&& A_{2,k} \equiv -6 (\pb+2)(\pb+3) w_{\pb+1;2(\pb+1),k}.
\end{eqnarray*}
Hence, the solutions extend analytically
through the sets $\mathcal{I}^\pm$, consistently with the theorem
concerning the regularity condition
(\ref{Helmut:regularity:condition}).

\subsection{Properties of the transport equations for $p=\pb+2$}
\label{section:pp2}

The solutions for $\mathfrak{S}_{\pb+1}[\upsilon^{(\pb+2)}]$ can be
calculated using the solutions for
$\mathfrak{S}_{\pb+1}[\phi^{(\pb+1)}]$ given by
(\ref{Weyl:pp1_0})-(\ref{Weyl:pp1_4}). According to lemma
\ref{lemma:logs:lead:to:logs} the solutions for
$\mathfrak{S}_{\pb+1}[\upsilon^{(\pb+2)}]$ will be polynomial in
$\tau$ and it turns out that they can be expressed in terms of
hypergeometric functions. With this information at hand it would be,
in principle, possible to analyse the Maxwell-like
reduced system (\ref{reduced:mp1})-(\ref{reduced:mp3}) order $p=\pb+2$ for the
coefficients $a_{1,\pb+2}$ and $a_{3,\pb+2}$. However, the form of the
explicit solutions in $\mathfrak{S}_{\pb+1}[\upsilon^{(\pb+2)}]$ make
it very difficult to identify useful structures in the
equations. Instead, it is desirable to find a way to extract
information on the solutions to the $p=\pb+2$ Maxwell-like reduced
system without having to make use of the explicit solutions of
$\mathfrak{S}_{\pb+1}[\upsilon^{(\pb+2)}]$.

\medskip
In order to get around this problem we do the following: a close
inspection of the right hand side of equations (\ref{reduced:mp1})
and (\ref{reduced:mp3}) reveals that by differentiating three times with
respect to $\tau$ and then using the $\upsilon^{(\pb+2)}$ and
$\phi^{(\pb+1)}$ transport equations and $\tau$-derivatives thereof,
it is possible to obtain a reduced system for
\[
 a_{1,\pb+2}^{[3]}\equiv \partial^3_\tau  a_{1,\pb+2}, \quad  a_{3,\pb+2}^{[3]}\equiv  \partial^3_\tau a_{3,\pb+2},
\]
where the non-homogeneous terms depend explicitly only on
$a_{1,\pb+1}$ and $a_{3,\pb+1}$ as given by expressions
(\ref{Weyl:pp1_1}) and (\ref{Weyl:pp1_3}). Clearly, if the solutions
$a_{1,\pb+2}^{[3]}$ and $ a_{3,\pb+2}^{[3]}$ for these equations are
regular at $\tau=\pm 1$, this will also be the case for $a_{1,\pb+2}$
and $ a_{3,\pb+2}$. Similarly, if $a_{1,\pb+2}^{[3]}$ and $
a_{3,\pb+2}^{[3]}$ have logarithmic singularities at $\tau=\pm 1$,
also $a_{1,\pb+2}$ and $ a_{3,\pb+2}$ will also contain them.

\medskip
The calculation of suitable equations for $a_{1,\pb+2}^{[3]}$ and $
a_{3,\pb+2}^{[3]}$ requires systematic and extensive use of computer
algebra methods. Calculations in the system {\tt Maple V} render a
reduced system of the form:
\begin{subequations}
\begin{eqnarray}
&& (1+\tau)\partial_\tau a_{1,\pb+2}^{[3]} + \frac{1}{2(\pb-1)}\bigg( (\pb+1)(\pb+2)\tau -(\pb^2-9\pb+2)  \bigg) a_{1,\pb+2}^{[3]} \nonumber \\
&& \hspace{1cm}+ \frac{1}{2(\pb-1)}(\pb+1)(\pb+2)(1-\tau) a_{3,\pb+2}^{[3]} \nonumber\\
&& \hspace{2cm}= \frac{1}{(1-\tau)^3(1+\tau)^4}G_{\pb+2} a_{1,\pb+1} +  \frac{1}{(1-\tau)^3(1+\tau)^4}H_{\pb+2}a_{3,\pb+1},\label{reducedpp2a}\\
&& (1-\tau)\partial_\tau a_{3,\pb+2}^{[3]} -\frac{1}{2(\pb-1)} (\pb+1)(\pb+2)(1+\tau)a^{[3]}_{1,\pb+2} \nonumber \\
&& \hspace{1cm}+ \frac{1}{2(\pb-1)}\bigg( (\pb+1)(\pb+2)\tau +(\pb^2-9\pb+2)  \bigg) a_{3,\pb+2}^{[3]} \nonumber \\
&& \hspace{2cm} =  \frac{1}{(1+\tau)^3(1-\tau)^4}H^s_{\pb+2}a_{1,\pb+1} +  \frac{1}{(1+\tau)^3(1-\tau)^4}G^s_{\pb+2}a_{3,\pb+1}. \label{reducedpp2b}
\end{eqnarray}
\end{subequations}
In the above expressions $G_{\pb+2}$ and $H_{\pb+2}$ are explicit
polynomials in $\tau$ of degree 9 with coefficients which are
themselves polynomials in $\pb$. Furthermore, $G^s_{\pb+2}(\tau)\equiv
G_{\pb+2}(-\tau)$, $H^s_{\pb+2}(\tau)\equiv H_{\pb+2}(-\tau)$. Both
$G_{\pb+2}$ and $H_{\pb+2}$ contain an overall factor of $m$. The
explicit form of these polynomials is given in the appendix. The
system (\ref{reducedpp2a})-(\ref{reducedpp2b}) is supplemented by the
initial conditions
\begin{equation}
\label{reducedpp2:data}
a^{[3]}_{1,\pb+2}(0)=-2mA_1(6\pb+23)(\pb+2), \quad a^{[3]}_{3,\pb+2}(0)=2mA_1(6\pb+23)(\pb+2).
\end{equation}
The data is calculated by using the values of
$\phi^{(\pb+1)}(0)$, $\phi^{(\pb+2)}(0)$, $\upsilon^{(\pb+2)}(0)$
implied by the data jets $J_{\mathcal{I}}^{(\pb+2)}[\upsilon]$,
$J_{\mathcal{I}}^{(\pb+2)}[\phi]$ and by $\tau$-differentiating as
necessary the $\upsilon^{(\pb+2)}$ and $\phi^{(\pb+2)}$ transport
equations. This cumbersome calculation has been carried out in the
computer algebra system {\tt Maple V}.

\medskip
Substitution of the explicit $\phi^{(\pb+1)}$ solutions
(\ref{Weyl:pp1_0})-(\ref{Weyl:pp1_4}) into equations
(\ref{reducedpp2a}) and (\ref{reducedpp2b}) shows that the right hand
sides of these equations are, respectively, of the form
\[
(1-\tau)^{\pb-3}(1+\tau)^{\pb-4} Q_{\pb+2}(\tau), \quad (1+\tau)^{\pb-3}(1-\tau)^{\pb-4} Q_{\pb+2}^s(\tau),
\]
where $Q_{\pb+2}(\tau)$ is a polynomial of degree 11 in $\tau$ such
that $Q_{\pb+2}(\pm 1)\neq 0$. Consistently with the above, it is
assumed that $\pb\geq 4$. The cases with $\pb<4$ can be analysed in a case by case basis ---cfr. the calculations in \cite{Val04a}.

\medskip
The remarkable structure of the zeros of the right hand sides of equations
(\ref{reducedpp2a}) and (\ref{reducedpp2b}) eases the task of looking
for polynomial solutions to these equations. Indeed, we note the
following lemma.

\begin{lemma}
\label{lemma:Galois:2}
All polynomial solutions of equations (\ref{reducedpp2a}) and
(\ref{reducedpp2b}) for $\pb\geq4$ are of the form
\begin{subequations}
\begin{eqnarray}
&& a_{1,\pb+2}^{[3]}(\tau)= (1-\tau)^{\pb-2} (1+\tau)^{\pb-4} b_{\pb+2}(\tau), \label{a1_polynomial_pp2}\\
&& a_{3,\pb+2}^{[3]}(\tau)= -(1+\tau)^{\pb-2} (1-\tau)^{\pb-4} b^s_{\pb+2}(\tau),\label{a3_polynomial_pp2} 
\end{eqnarray}
\end{subequations}
where $b_{\pb+2}(\tau)$ is polynomial of degree 9 and $b^s_{\pb+2}(\tau)\equiv b_{\pb+2}(-\tau)$.  
\end{lemma}

\noindent
\textbf{Proof.} The proof of this lemma is inspired by the discussion
in chapter 4 of \cite{PutSin03}. One starts by looking at the possible zeros
of the solution to equations (\ref{reducedpp2a}) and
(\ref{reducedpp2b}) at $\tau=1$.  If $a_1^{[3]}$ and $a_3^{[3]}$ are
polynomial then one can write
\begin{equation}
\label{Ansatz:polynomial}
a_1^{[3]}= \sum_{k=n_*}^{n^*} \alpha_k (1-\tau)^k, \quad a_3^{[3]}= \sum_{k=m_*}^{m^*} \beta_k (1-\tau)^k,
\end{equation}
for some integers $n_*, \; m_*,\; n^*, \;m^*\geq 0$ and some complex
numbers $\alpha_k$, $\beta_l$, $n_*\leq k \leq n^*$, $m_* \leq l \leq
m^*$. Dividing equation (\ref{reducedpp2a}) by $1+\tau$ and equation
(\ref{reducedpp2b}) by $1-\tau$ and then substituting the expressions
(\ref{Ansatz:polynomial}) into the ordinary differential equations
(\ref{reducedpp2a}) and (\ref{reducedpp2b}) one finds that
\[
\min \{n_*-1, m_*+1  \} =\pb-3, \quad \min \{n_*-1, m_*-1 \}=\pb-5,
\]
as $\tau=1$ is a zero of the right hand sides of (\ref{reducedpp2a})
and (\ref{reducedpp2b}) and, moreover, the left and right hand sides
must have the same multiplicity. The above conditions are satisfied by
setting
\[
n_*=\pb-2, \quad m_*=\pb-4.
\]
The discussion of zeros at $\tau=-1$ follows by symmetry ---cfr. the
discussion in section \ref{section:discrete:symms}. The degree of the
polynomial $b_{\pb+2}(\tau)$ follows by inspection of
$Q_{\pb+2}$. \hfill $\Box$

\bigskip
Now, the substitution of the Ansatz (\ref{a1_polynomial_pp2}) and (\ref{a3_polynomial_pp2}) with
\[
b_{\pb+2}(\tau) = \sum_{k=0}^{9} B_{\pb+2,k} \tau^k, \quad B_{\pb+2,k}\in\Complex
\]
into the equations (\ref{reducedpp2a}) and (\ref{reducedpp2b}) 
leads to a system of 11 linear algebraic equations for the 10 unknowns
$B_{\pb+2,k}$, $k=0,\ldots,9$. This overdetermined system can be seen to have
a solution which can be explicitly calculated. The result of this
calculation is also presented in the appendix. In particular, the solution
obtained is such that
\[
a^{[3]}_{1,\pb+2}(0)=-2 m A_1 (6\pb+23)(\pb+2), \quad a^{[3]}_{3,\pb+2}(0)=2 m A_1 (6\pb+23)(\pb+2),
\]
consistent with the initial data (\ref{reducedpp2:data}) for the
system (\ref{a1_polynomial_pp2}) and (\ref{a3_polynomial_pp2}). Thus,
the particular solution to the system
(\ref{reducedpp2a})-(\ref{reducedpp2b}) calculated by the above
procedure is, in fact, the solution to the system in question with data
given by (\ref{reducedpp2:data}).

\medskip
From the procedure described in the above paragraphs together with explicit calculations in \cite{Val04a} for $\pb<4$ one has the
following result.

\begin{proposition}
The solution of the Maxwell-like reduced system (\ref{reducedpp2a})
and (\ref{reducedpp2b}) with data given by (\ref{reducedpp2:data}) is
polynomial in $\tau$ and thus it extends smoothly through $\tau=\pm1$.
\end{proposition}

\medskip
\noindent
\textbf{Remark.} It follows that not only
$a^{[3]}_{1,\pb+2}$ and $a^{[3]}_{3,\pb+2}$
but also $a_{1,\pb+2}$ and $a_{3,\pb+2}$ are
polynomial in $\tau$. Moreover, from the discussion in section
\ref{section:alternative:reduced} and lemma \ref{lemma:Galois:2} also
$a_{0,\pb+2}$, $a_{2,\pb+2}$,
$a_{4,\pb+2}$ and the sector
$\mathfrak{S}_{\pb+1}[\upsilon^{(\pb+3)}]$ are polynomial in $\tau$.

\subsection{Properties of the transport equations for $p=\pb+3$}
\label{section:pp3}

We extend the approach used in the previous section to decide
whether the coefficients $a_{1,\pb+3}$ and $a_{3,\pb+3}$ are
polynomial in $\tau$ or not. For these coefficients, the calculations
are more involved, and require $\tau$-differentiating the
Maxwell-like reduced equations satisfied by $a_{1,\pb+3}$ and
$a_{3,\pb+3}$ seven times to obtain a linear system for
\[
 a_{1,\pb+3}^{[7]}\equiv \partial^7_\tau  a_{1,\pb+3}, \quad  a_{3,\pb+3}^{[7]}\equiv  \partial^7_\tau a_{3,\pb+3}.
\]
After computer algebra computations involving the substitution
of the transport equations satisfied by $\upsilon^{(\pb+2)}$,
$\upsilon^{(\pb+3)}$ and $\phi^{(\pb+1)}$, $\phi^{(\pb+2)}$ and
$\tau$-derivatives thereof, one obtains the following system of
equations:
\begin{subequations}
\begin{eqnarray}
&& (1+\tau)\partial_\tau a_{1,\pb+3}^{[7]} - \frac{1}{2(\pb-4)}\bigg( (\pb+1)(\pb+2)\tau -(\pb^2-21\pb-38)  \bigg) a_{1,\pb+3}^{[7]} \nonumber \\
&& \hspace{1cm} - \frac{1}{2(\pb-4)}(\pb+1)(\pb+2)(1-\tau) a_{3,\pb+3}^{[7]} \nonumber \\
&& \hspace{2cm}= \frac{1}{(1-\tau)^8(1+\tau)^9}G_{\pb+3} a_{1,\pb+1} +  \frac{1}{(1-\tau)^8(1+\tau)^9}H_{\pb+3}a_{3,\pb+1}\nonumber \\
&& \hspace{3cm}+ \frac{1}{(1-\tau)^4(1+\tau)^5}K_{\pb+3} a^{[3]}_{1,\pb+2} +  \frac{1}{(1-\tau)^4(1+\tau)^5}L_{\pb+3}a^{[3]}_{3,\pb+2}, \nonumber \\
&& \label{reducedpp3a}\\
&& (1-\tau)\partial_\tau a_{3,\pb+3}^{[7]} +\frac{1}{2(\pb-4)} (\pb+1)(\pb+2)(1+\tau)a^{[7]}_{1,\pb+3} \nonumber \\
&& \hspace{1cm} - \frac{1}{2(\pb-4)}\bigg( (\pb+1)(\pb+2)\tau +(\pb^2-21\pb+38)  \bigg) a_{3,\pb+3}^{[3]} \nonumber \\
&& \hspace{2cm} =  \frac{1}{(1+\tau)^8(1-\tau)^9}H^s_{\pb+3}a_{1,\pb+1} +  \frac{1}{(1+\tau)^8(1-\tau)^9}G^s_{\pb+3}a_{3,\pb+1} \nonumber \\
&& \hspace{3cm}   +\frac{1}{(1+\tau)^4(1-\tau)^5}L^s_{\pb+3}a^{[3]}_{1,\pb+2} +  \frac{1}{(1+\tau)^4(1-\tau)^5}K^s_{\pb+3}a^{[3]}_{3,\pb+2}. \nonumber \\
&& \label{reducedpp3b}
\end{eqnarray}
\end{subequations}

In the above expressions $G_{\pb+3}$ and $H_{\pb+3}$ are explicit
polynomials in $\tau$ of degree 19, while $K_{\pb+3}$ and $L_{\pb+3}$
are explicit polynomials of degree 10. The polynomials $G_{\pb+3}$ and
$H_{\pb+3}$ contain an overall factor of $m^2$ while $K_{\pb+3}$ and
$L_{\pb+3}$ have an overall factor of $m$. The coefficients of these
polynomials are themselves polynomials in $\pb$. As in the case of $p=\pb+2$, initial conditions for the system (\ref{reducedpp3a})-(\ref{reducedpp3b}) are calculated using the values of
$\phi^{(\pb+1)}(0)$, $\phi^{(\pb+2)}(0)$, $\phi^{(\pb+3)}(0)$, $\upsilon^{(\pb+2)}(0)$, $\upsilon^{(\pb+3)}(0)$ 
implied by the data jets $J_{\mathcal{I}}^{(\pb+2)}[\upsilon]$, $J_{\mathcal{I}}^{(\pb+3)}[\upsilon]$, 
$J_{\mathcal{I}}^{(\pb+2)}[\phi]$, $J_{\mathcal{I}}^{(\pb+3)}[\phi]$  and by $\tau$-differentiating as
necessary the $\upsilon^{(\pb+3)}$ and $\phi^{(\pb+3)}$ transport
equations. One obtains
\begin{subequations}
\begin{eqnarray}
&& a^{[7]}_{1,\pb+7}(0)=\frac{1}{8}m^2A(\pb+2)(\pb+3)(3\pb^5 +72\pb^4 +49353\pb^3  \nonumber \\
&& \hspace{6cm}+ 142260\pb^2 + 610272\pb + 302048), \label{reducedpp3a:data}\\ 
&& a^{[7]}_{3,\pb+7}(0)=-\frac{1}{8}m^2A(\pb+2)(\pb+3)(3\pb^5 +72\pb^4 +49353\pb^3 \nonumber \\
&& \hspace{6cm}+ 142260\pb^2 + 610272\pb + 302048). \label{reducedpp3b:data}
\end{eqnarray}
\end{subequations}

\bigskip
We proceed to analyse this system on the same lines as it was done for
(\ref{reducedpp2a})-(\ref{reducedpp2b}). First one notices that
direct substitution of the explicit solutions $a_{1,\pb+1}$,
$a_{3,\pb+1}$, $a_{1,\pb+2}$ and $a_{3,\pb+2}$ which were obtained,
respectively, in sections \ref{section:pp1} and \ref{section:pp2} shows
that the right hand sides of equations (\ref{reducedpp3a}) and
(\ref{reducedpp3b}) are of the form
\[
(1-\tau)^{\pb-8}(1+\tau)^{\pb-9} Q_{\pb+3}(\tau), \quad (1+\tau)^{\pb-8}(1-\tau)^{\pb-9} Q^s_{\pb+3}(\tau)
\]
where $Q_{\pb+3}(\tau)$ is a polynomial in $\tau$ such that
$Q_{\pb+3}(\pm 1)\neq 0$. In order for the following calculations to
make sense, it is assumed that $\pb\geq 9$. The cases $\pb<9$ can be
analysed individually and provide the same qualitative picture.

\medskip
The remarkable structure of the zeros of the right hand sides of equations
(\ref{reducedpp2a}) and (\ref{reducedpp2b}) eases the task of looking
for polynomial solutions to these equations. Indeed, one has the
following lemma, whose proof is similar to that of lemma \ref{lemma:Galois:2}.

\begin{lemma}
\label{lemma:Galois:3}
All polynomial solutions of equations (\ref{reducedpp3a}) and
(\ref{reducedpp3b}) are of the form
\begin{subequations}
\begin{eqnarray}
&& a_{1,\pb+3}^{[7]}(\tau)= (1-\tau)^{\pb-7} (1+\tau)^{\pb-9} b_{\pb+3}(\tau), \label{a1_polynomial_pp3}\\
&& a_{3,\pb+3}^{[7]}(\tau)= -(1+\tau)^{\pb-7} (1-\tau)^{\pb-9} b^s_{\pb+3}(\tau),\label{a3_polynomial_pp3} 
\end{eqnarray}
\end{subequations}
where $b_{\pb+3}(\tau)$ is polynomial of degree 19, and
$b^s_{\pb+3}(\tau)\equiv b_{\pb+3}(-\tau)$.
\end{lemma}

\bigskip
Now, we use the expressions (\ref{a1_polynomial_pp3}) and (\ref{a3_polynomial_pp3}) as an Ansatz for $a^{[7]}_{1,\pb+3}$ and $a^{[7]}_{3,\pb+3}$ with
\[
b_{\pb+3}(\tau) = \sum_{k=0}^{19} B_{\pb+3,k} \tau^k, \quad B_{\pb+3,k}\in \Complex.
\]
The substitution of this Ansatz into equations
(\ref{reducedpp3a})-(\ref{reducedpp3b}) leads to a system of 21 linear
algebraic equations for the 20 unknowns $B_{\pb+3,k}$,
$k=0,\ldots,19$. By means of an explicit calculation, this
overdetermined system can be seen to have a solution ---that is, not
all the 21 equations are linearly independent. The solution so
obtained contains a free parameter which can be adjusted to set the
value of the coefficient $B_{\pb+3,0}$ such that the initial
conditions (\ref{reducedpp3a:data})-(\ref{reducedpp3b:data}) are
satisfied.

\medskip
With the procedure described in the above paragraphs and explicit calculations for the cases with $\pb<9$ one has the
following result.

\begin{proposition}
The solution of the Maxwell-like reduced system
(\ref{reducedpp3a})-(\ref{reducedpp3b}) with data given by
(\ref{reducedpp3a:data})-(\ref{reducedpp3b:data}) is polynomial in
$\tau$ and thus, it extends smoothly through $\tau=\pm1$.
\end{proposition}

\medskip
\noindent
\textbf{Remark.} It follows that not only $a^{[7]}_{1,\pb+3}$ and
$a^{[7]}_{3,\pb+3}$ but also $a_{1,\pb+3}$ and $a_{3,\pb+3}$ are
polynomial in $\tau$. Moreover, from the discussion in section
\ref{section:alternative:reduced} and lemma \ref{lemma:Galois:3},
$a_{0,\pb+3}$ $a_{2,\pb+3}$, $a_{4,\pb+3}$ and the sector
$\mathfrak{S}_{\pb+1}[\upsilon^{(\pb+4)}]$ are polynomial in $\tau$.

\subsection{Properties of the transport equations for $p=\pb+4$}
\label{section:pp4}
 
Finally, it is shown, by methods similar to the ones used for the
cases $p=\pb+2$ and $p=\pb+3$ that the solutions $a_{1,\pb+4}$ and
$a_{3,\pb+4}$ of the order $p=\pb+4$ Maxwell-like transport
equations cannot be purely polynomial. In order to show this, one
$\tau$-differentiates eleven times(!) the Maxwell-like reduced equations
satisfied by $a_{1,\pb+4}$ and $a_{3,\pb+4}$ to obtain a linear system
for
\[
a^{[11]}_{1,\pb+4} \equiv \partial^{11}_\tau a_{1,\pb+4}, \quad a^{[11]}_{3,\pb+4}\equiv \partial^{11}_\tau a_{3,\pb+4}.
\]
After lengthy computer algebra computations involving the substitution
of the transport equations satisfied by $\upsilon^{\pb+4}$,
$\upsilon^{\pb+3}$, $\upsilon^{\pb+2}$, $\phi^{\pb+3}$,
$\phi^{\pb+2}$, $\phi^{\pb+1}$ and $\tau$-derivatives thereof into the
equations for $a^{[11]}_{1,\pb+4}$ and $a^{[11]}_{3,\pb+4}$, one
obtains a system of equations of the form:
\begin{subequations}
\begin{eqnarray}
&& (1+\tau)\partial_\tau a_{1,\pb+4}^{[11]} - \frac{1}{2(\pb-7)}\bigg( (\pb+1)(\pb+2)\tau -(\pb^2-33\pb+110)  \bigg) a_{1,\pb+4}^{[11]} \nonumber \\
&& \hspace{1cm} - \frac{1}{2(\pb-7)}(\pb+1)(\pb+2)(1-\tau) a_{3,\pb+4}^{[11]} \nonumber \\
&& \hspace{2cm}= \frac{1}{(1-\tau)^{13}(1+\tau)^{14}}G_{\pb+4} a_{1,\pb+1} +  \frac{1}{(1-\tau)^{13}(1+\tau)^{14}}H_{\pb+4}a_{3,\pb+1} \nonumber \\
&& \hspace{3cm}+ \frac{1}{(1-\tau)^9(1+\tau)^{10}}K_{\pb+4} a^{[3]}_{1,\pb+2} +  \frac{1}{(1-\tau)^9(1+\tau)^{10}}L_{\pb+4}a^{[3]}_{3,\pb+2} \nonumber \\
&& \hspace{4cm} + \frac{1}{(1-\tau)^4(1+\tau)^5}M_{\pb+4} a^{[7]}_{1,\pb+3} +  \frac{1}{(1-\tau)^4(1+\tau)^5}N_{\pb+4}a^{[7]}_{3,\pb+3}, \nonumber \\
&& \label{reducedpp4a}\\
&& (1-\tau)\partial_\tau a_{3,\pb+4}^{[11]} +\frac{1}{2(\pb-7)} (\pb+1)(\pb+2)(1+\tau)a^{[11]}_{1,\pb+4} \nonumber \\
&& \hspace{1cm} - \frac{1}{2(\pb-7)}\bigg( (\pb+1)(\pb+2)\tau +(\pb^2-33\pb+110)  \bigg) a_{3,\pb+4}^{[11]} \nonumber \\
&& \hspace{2cm} =  \frac{1}{(1+\tau)^{13}(1-\tau)^{14}}H^s_{\pb+4}a_{1,\pb+1} +  \frac{1}{(1+\tau)^{13}(1-\tau)^{14}}G^s_{\pb+4}a_{3,\pb+1} \nonumber \\
&& \hspace{3cm}   +\frac{1}{(1+\tau)^9(1-\tau)^{10}}L^s_{\pb+4}a^{[3]}_{1,\pb+2} +  \frac{1}{(1+\tau)^9(1-\tau)^{10}}K^s_{\pb+4}a^{[3]}_{3,\pb+2} \nonumber \\
&& \hspace{4cm} + \frac{1}{(1+\tau)^4(1-\tau)^5}N^s_{\pb+4} a^{[7]}_{1,\pb+3} +  \frac{1}{(1+\tau)^4(1-\tau)^5}M^s_{\pb+4}a^{[7]}_{3,\pb+3}. \nonumber \\
&& \label{reducedpp4b}
\end{eqnarray}
\end{subequations}
In the above expressions $G_{\pb+4}$ and $H_{\pb+4}$ are polynomials
in $\tau$ of degree 29 with an overall factor of $m^3$; $K_{\pb+4}$
and $L_{\pb+4}$ are polynomials in $\tau$ of degree 20 with an overall
factor of $m^2$; finally, $M_{\pb+4}$ and $N_{\pb+4}$ are polynomials
in $\tau$ of degree 9 with an overall factor of $m$. The coefficients
of all these polynomials are themselves polynomial in $\pb$. The
initial conditions for the system
(\ref{reducedpp4a})-(\ref{reducedpp4b}) can be calculated following
a similar procedure to the one used for the lower order systems
(\ref{reducedpp2a})-(\ref{reducedpp2b}) and
(\ref{reducedpp3a})-(\ref{reducedpp3b}). The initial conditions read
\begin{eqnarray}
&& a_{1,\pb+4}^{[11]}(0)=-\frac{1}{8}m^3A(\pb+2)(\pb+3)(\pb+4)(1737\pb^7+64317\pb^6 +84931161\pb^5 -175033869\pb^4 \nonumber \\
&& \hspace{3cm}+ 3959957166\pb^3-1379005296\pb^2 +6565457856\pb+1464602048) \label{reducedpp4a:data}\\
&&  a_{3,\pb+4}^{[11]}(0)=\frac{1}{8}m^3A(\pb+2)(\pb+3)(\pb+4)(1737\pb^7+64317\pb^6 +84931161\pb^5 -175033869\pb^4 \nonumber \\
&& \hspace{3cm}+ 3959957166\pb^3-1379005296\pb^2 +6565457856\pb+1464602048). \label{reducedpp4b:data}
\end{eqnarray}

\bigskip
As in the previous cases, the crucial observation is that using the
explicit expressions for $a_{1,\pb+1}$, $a_{3,\pb+1}$,
$a^{[3]}_{1,\pb+2}$, $a^{[3]}_{3,\pb+2}$, $a^{[7]}_{1,\pb+3}$ and
$a^{[7]}_{3,\pb+3}$ discussed in the previous sections, one finds that
the right hand sides of equations (\ref{reducedpp4a}) and
(\ref{reducedpp4b}) are, respectively, of the form
\[
(1-\tau)^{\pb-13}(1+\tau)^{\pb-14}Q_{\pb+4}(\tau), \quad -(1+\tau)^{\pb-13}(1-\tau)^{\pb-14}Q^s_{\pb+4}(\tau),
\]
with $Q_{\pb+4}(\tau)$ a polynomial in $\tau$ such that
$Q_{\pb+4}(\pm1)\neq 0$. In what follows it is assumed that $\pb\geq
14$, so that the discussions make
sense. The cases $\pb<14$ can be dealt with direct
case-by-case calculations. The qualitative results are the same as in
the general case.

\medskip
 As in the cases discussed in sections
\ref{section:pp2} and \ref{section:pp3}, the structure of the
polynomial solutions (if any) of the system
(\ref{reducedpp4a})-(\ref{reducedpp4b}) is very particular. More
precisely, one has the following result.

\begin{lemma}
The polynomial solutions (if any) of the reduced system
(\ref{reducedpp4a}) and (\ref{reducedpp4b}) are of the form
\begin{subequations}
\begin{eqnarray}
&& a_{1,\pb+4}^{[11]}(\tau)= (1-\tau)^{\pb-12} (1+\tau)^{\pb-14} b_{\pb+4}(\tau), \label{a1_polynomial_pp4}\\
&& a_{3,\pb+4}^{[11]}(\tau)= -(1+\tau)^{\pb-12} (1-\tau)^{\pb-14} b^s_{\pb+4}(\tau),\label{a3_polynomial_pp4} 
\end{eqnarray}
\end{subequations}
where $b_{\pb+4}$ is a polynomial of degree 29.
\end{lemma}

\bigskip
As in the previous orders, we use expressions
(\ref{a1_polynomial_pp4})-(\ref{a3_polynomial_pp4}) as an Ansatz
for $a^{[11]}_{1,\pb+4}$ and $a^{[11]}_{3,\pb+4}$ with
\[
b_{\pb+4}(\tau) = \sum_{k=0}^{29} B_{\pb+4,k} \tau^k, \quad B_{\pb+4,k}\in \Complex.
\]
Again, this Ansatz leads to a system of 31 linear algebraic equations
for the 30 unknowns $B_{\pb+4,k}$, $k=0,\ldots,29$. By means of an
explicit calculation, this overdetermined system can be seen to have
no solutions unless $w_{\pb+1,2\pb+2,k}=0$ or $m=0$. This is the
crucial result of our analysis. One has the following proposition.

\begin{proposition}
The solution of the Maxwell-like reduced system (\ref{reducedpp4a})
and (\ref{reducedpp4b}) admits no polynomial solutions unless
$w_{\pb+1,2\pb+2,k}=0$ or $m=0$.
\end{proposition}

This last proposition, together with proposition
\ref{proposition:logarithmic:solutions} in section
\ref{section:general:properties:maxwell} renders the following
corollary.

\begin{corollary}
The solution of the Maxwell-like reduced system
(\ref{reducedpp4a})-(\ref{reducedpp4b}) with data given by
(\ref{reducedpp4a:data})-(\ref{reducedpp4b:data}) develops logarithmic
singularities at $\tau=\pm1$ unless $w_{\pb+1,2\pb+2,k}=0$ or
$m=0$. The logarithmic solutions are of class $C^\omega(-1,1)\cap
C^{\pb+3}[-1,1]$.
\end{corollary}

\medskip
\noindent
\textbf{Remark.} From the discussion in section
\ref{section:alternative:reduced} it follows that the coefficients
$a_{i,\pb+4}$, $i=0,\ldots,4$ will have logarithmic singularities of
the form given by proposition \ref{proposition:logarithmic:solutions}.

\section{The main result}
\label{section:main}

The discussion in the previous sections is summarised in the following result.

\begin{proposition}
  Given a time symmetric initial data set which in a neighbourhood
  $\mathcal{B}_a$ of infinity is Schwarzschildean up to order $p=\pb$,
  the solutions to the transport equations for the orders $p=\pb+1$,
  $\pb+2$, $\pb+3$ are polynomial in $\tau$ and hence extend smoothly
  through the critical sets $\mathcal{I}^\pm$. On the other hand, the
  solutions at order $p=\pb+4$ contain logarithmic singularities which
  can be avoided if and only if the initial data is, in fact,
  Schwarzschildean to order $p=\pb+1$.
\end{proposition}

From this result, an induction argument which uses the explicit
calculations performed in \cite{Val04a} as a base step renders our
main result ---cfr. the main theorem in the introduction.

\begin{theorem}
\label{thm:main}
  The solution to the regular finite initial value problem at spatial
  infinity for initial data which is time symmetric and conformally flat in
  a neighbourhood of infinity is smooth through $\mathcal{I}^\pm$ if
  and only if the restriction of the data to $\mathcal{I}^0$ coincides
  with the restriction of Schwarzschildean data at every
  order. Furthermore, the analyticity of the data implies that the
  initial data is exactly Schwarzschildean in a neighbourhood of
  infinity.
\end{theorem}

As mentioned in the introduction, the evidence gathered in
\cite{Val04d} suggests that it is possible to obtain a generalisation
of this result for initial data sets which are not conformally
flat. In that case, the expected conclusion is that smoothness at the
critical sets implies staticity at every order. In \cite{Fri04} it has
been shown that the conformal structure of static spacetimes is as
regular as one would expect it to be. In particular, static spacetimes
extend smoothly (in fact analytically) at the critical sets
$\mathcal{I}^\pm$. Key to such a generalisation of theorem \ref{thm:main}
is to consider a class of time symmetric initial data sets for which
it is simple to decide whether its development will be a static
spacetime or not.

\section*{Acknowledgements}
This research is funded by an EPSRC Advanced Research Fellowship. I
thank Helmut Friedrich for introducing me to this problem many years
ago, and for many helpful discussions through the years. Thanks are
also due to Robert Beig, Malcolm MacCallum, Christian L\"ubbe, Thomas
B\"ackdahl and Jos\'e Luis Jaramillo for several discussions. I thank
ChristianeM Losert-VK for support, encouragement and a careful reading
of the manuscript.

\appendix

\section{Some of the polynomials in section \ref{section:punchline}}
In order to exemplify the type of expressions one has to deal with in
the analysis described in section \ref{section:punchline}, here we
present some of the simpler polynomials.

\subsection{The polynomials in section \ref{section:pp2}}
{\footnotesize
\begin{eqnarray*}
&&\frac{1}{m(\pb+2)}G_{\pb+2}=\left( \pb+4 \right)  \left( \pb+2 \right)  \left( 2\,{\pb}^{3}+10\,{\pb}^{2}
-2\,\pb-55 \right) {\tau}^{9}- \left( 17\,{\pb}^{3}+635+305\,\pb+14\,{\pb}^{4}
+2\,{\pb}^{5}+8\,{\pb}^{2} \right) {\tau}^{8}\\
&&-2\, \left( 175\,{\pb}^{2}+7\,{
\pb}^{5}-550\,\pb+246\,{\pb}^{3}-652+72\,{\pb}^{4} \right) {\tau}^{7}-2\,
 \left( 5\,{\pb}^{5}+155\,{\pb}^{2}+115\,{\pb}^{3}+34\,{\pb}^{4}-471\,\pb-1080
 \right) {\tau}^{6}+ \\
&& \left( 833\,{\pb}^{3}-1124+795\,{\pb}^{2}+12\,{\pb}^{5}
-1172\,\pb+206\,{\pb}^{4} \right) {\tau}^{5}+ \left( 36\,{\pb}^{5}-2494+601
\,{\pb}^{3}-946\,\pb+166\,{\pb}^{4}+891\,{\pb}^{2} \right) {\tau}^{4} \\
&&+2\,
 \left( 263\,\pb+70+12\,{\pb}^{4}-280\,{\pb}^{2}-263\,{\pb}^{3} \right) {\tau}
^{3}-12\, \left( -32\,\pb-95+78\,{\pb}^{2}+12\,{\pb}^{4}+22\,{\pb}^{3}
 \right) {\tau}^{2} \\
&&-3\, \left( -62\,{\pb}^{2}-72+51\,{\pb}^{3}+44\,\pb
 \right) \tau-75+45\,\pb+54\,{\pb}^{3}+39\,{\pb}^{2},
\end{eqnarray*}

\begin{eqnarray*}
&& \frac{1}{m(\pb+2)}H_{\pb+2}=-2\, \left( \pb+4 \right)  \left( \pb+2 \right)  \left( 2\,{\pb}^{3}+10\,{\pb}
^{2}-2\,\pb-55 \right) {\tau}^{9}+2\, \left( 161\,\pb+102\,{\pb}^{2}+18\,{\pb}
^{4}+33+53\,{\pb}^{3}+2\,{\pb}^{5} \right) {\tau}^{8}\\
&&+4\, \left( \pb+2
 \right)  \left( 7\,{\pb}^{4}+56\,{\pb}^{3}+122\,{\pb}^{2}-71\,\pb-450
 \right) {\tau}^{7}+4\, \left( 3\,{\pb}^{3}-113\,{\pb}^{2}-305\,\pb+20\,{\pb}^
{4}-78+5\,{\pb}^{5} \right) {\tau}^{6}\\
&&-2\, \left( -1352\,\pb+905\,{\pb}^{3}+
959\,{\pb}^{2}-2748+12\,{\pb}^{5}+226\,{\pb}^{4} \right) {\tau}^{5}-2\,
 \left( 245\,{\pb}^{3}+36\,{\pb}^{5}+142\,{\pb}^{4}-866\,\pb-246-13\,{\pb}^{2}
 \right) {\tau}^{4}\\
&&+4\, \left( 335\,{\pb}^{3}+24\,{\pb}^{4}-359\,\pb+562\,{\pb
}^{2}-886 \right) {\tau}^{3}+48\, \left( 11\,{\pb}^{3}-7\,\pb-13+6\,{\pb}^{4
} \right) {\tau}^{2} \\
&&-6\, \left( -176+98\,{\pb}^{2}+21\,{\pb}^{3}+60\,\pb
 \right) \tau+90+78\,\pb-108\,{\pb}^{3}-186\,{\pb}^{2},
\end{eqnarray*}

\begin{eqnarray*}
&& \frac{1}{m A_1} b_{p+2} \left( \tau \right) =-2 \left( 6\,\pb+23 \right) 
 \left( \pb+2 \right) -2 \left( \pb+2 \right)  \left( 6\,{\pb}^{2}+14\,\pb
+5 \right) \tau+4\, \left( \pb+2 \right)  \left( 21\,{\pb}^{2}+45\,\pb+73
 \right){\tau}^{2}\\
&&+4 \left( \pb+2 \right)  \left( 7\,{\pb}^{3}+16\,
{\pb}^{2}+88\,\pb+48 \right) {\tau}^{3}-2 \left( \pb+2 \right)  \left( 
20\,{\pb}^{3}+55\,{\pb}^{2}+229\,\pb+256 \right) {\tau}^{4} \\
&&-2 \left( \pb+1
 \right)  \left( \pb+2 \right)  \left( 4\,{\pb}^{3}+15\,{\pb}^{2}+122\,\pb+214
 \right) {\tau}^{5} +\frac{4}{3} \left( \pb+2 \right)  \left( 7\,{\pb}^{3}+69\,
{\pb}^{2}+260\,\pb+249 \right) {\tau}^{6}\\
&&+\frac{4}{3} \left( {\pb}^{3}+29\,{\pb}^{
2}+127\,\pb+120 \right)  \left( \pb+2 \right) ^{2}{\tau}^{7} -\frac{2}{3}
 \left( \pb+2 \right)  \left( 2\,{\pb}^{3}+27\,{\pb}^{2}+109\,\pb+99 \right) {
\tau}^{8}\\
&& -\frac{2}{3}\, \left( \pb+2 \right)  \left( 2\,{\pb}^{4}+23\,{\pb}^{3}+97\,
{\pb}^{2}+178\,\pb+111 \right) {\tau}^{9}.
\end{eqnarray*}

}

The expressions of the polynomials appearing in the analyses at order
$p=\pb+3$ and $p=\pb+4$ are too unwieldly to be presented here.

% Path in Luise
%\bibliography{/home/jav/tex/grbib}
% Path in QM 
%\bibliography{/home/network/jav/tex/grbib}
%\bibliography{/home/jav/QM/tex/grbib}
% Path in Ludovica
%\bibliography{/Users/Juan/Documents/tex/grbib}

\end{document}